\documentclass[fleqn,usenatbib]{mnras}

\usepackage{newtxtext,newtxmath}

\usepackage[T1]{fontenc}

\DeclareRobustCommand{\DA}[3]{#2}
\let\DAthebibliography\thebibliography
\def\thebibliography{\DeclareRobustCommand{\DA}[3]{##3}\DAthebibliography}
\DeclareRobustCommand{\daname}[3]{#2}
\let\dathebibliography\thebibliography
\def\thebibliography{\DeclareRobustCommand{\daname}[3]{##3}\dathebibliography}

\usepackage{graphicx}	
\usepackage{amsmath}	

\newcommand{\bagpipes}{{\sc bagpipes}}

\title[SED Modelling Uncertainties]{Impact of Uncertainties in Spectral Energy Distribution Modelling on Inferred Galaxy Properties}

\author[G. T. Jones, C. M. Byrne, E. R. Stanway]{
Gareth T. Jones,$^{1}$\thanks{E-mail: g.jones.6@warwick.ac.uk}
Conor M. Byrne,$^{1}$
Elizabeth R. Stanway$^{1}$
\\
$^{1}$Department of Physics, University of Warwick, Gibbet Hill Road, Coventry CV4 7AL, UK
}

\date{Accepted 2025 August 28. Received 2025 August 21; in original form 2025 May 22}

\pubyear{\the\year{}}

\begin{document}
\label{firstpage}
\pagerange{\pageref{firstpage}--\pageref{lastpage}}
\maketitle

\begin{abstract}
Interpreting galaxy properties from astronomical surveys relies heavily on spectral energy distribution (SED) modelling, yet uncertainties in key model ingredients are often overlooked.
By leveraging a $z\sim0$ galaxy sample from the EAGLE simulation, we generate synthetic SDSS spectral and VISTA photometric observations with controlled assumptions, to assess how variations in stellar spectral library, initial mass function (IMF) and metallicity prescriptions within the BPASS-framework affect inferred galaxy properties.
Our analysis combines spectral fitting from 3800 to 9200\,\AA\ with photometric constraints extending to 2.3\,$\mu$m, enabling robust assessment across a broad wavelength baseline.
Our findings reveal mass, age and star formation rate vary by $0.27\pm0.09$, $0.19\pm0.11$ and $1.4\pm1.0$ dex, respectively, greater than observational uncertainties reported in surveys.
Notably, we find stellar spectral library choice is capable of transforming a galaxy from appearing star-forming to quiescent, while a fixed metallicity assumption yields systematic biases when the chosen metallicity is incorrect.
These modelling differences impact the reconstructed total mass assembly history in galaxies by up to $\sim12$ percent and bias the demographic and star formation history conclusions drawn from surveys.
As upcoming missions like \textit{Euclid}, \textit{Roman} and CASTOR aim to characterise galaxy evolution with unprecedented precision, our results highlight the need for careful propagation of SED modelling uncertainties and transparency in model selection.
\end{abstract}



\section{Introduction}
Astronomical surveys are a vital observational technique, enabling the simultaneous study of a large number of objects and generating vast datasets.
Numerous surveys have already been carried out (e.g. SDSS, \citealt{2000AJ....120.1579Y}, \citealt{2003MNRAS.341...33K}; COSMOS, \citealt{2016ApJS..224...24L}, \citealt{2022ApJS..258...11W}, \citealt{2023A&A...677A.184W}; Galaxy and Mass Assembly, \citealt{2011MNRAS.413..971D}, \citealt{2015MNRAS.452.2087L}, \citealt{2020MNRAS.498.5581B}; Cosmic Assembly Near-infrared Deep Extragalactic Legacy Survey, \citealt{2011ApJS..197...35G}, \citealt{2011ApJS..197...36K}, \citealt{2016MNRAS.462.4495H}), and more are forthcoming (e.g. CASTOR, \citealt{2012SPIE.8442E..15C, 2019clrp.2020...18C}; \textit{Roman}, \citealt{2015arXiv150303757S}, \citealt{2019arXiv190205569A}, \citealt{2022ApJ...928....1W}; \textit{Euclid}, \citealt{2024arXiv240513491E}), aiming to improve upon earlier efforts by delivering higher-quality data with reduced observational uncertainties.
However, physical properties cannot be directly extracted from these observations, and instead must be inferred using models and techniques such as spectral energy distribution (SED) fitting.
While widely used, SED fitting has its own systematic uncertainties, and there has been a recent surge in publications highlighting the substantial impact these uncertainties can have on the interpretation of derived galaxy properties \citep[e.g.][]{2019ApJ...876....3L, 2020ApJ...904...33L, 2022MNRAS.514.5706J, 2022ApJ...935..146S, 2022MNRAS.509.4940T, 2023MNRAS.525.5720J, 2024arXiv241017698B, 2024MNRAS.530L...7H, 2024ApJ...963...74W}.

SED fitting has become an essential tool that has been developed and refined over several decades since the early work by \citet{1968ApJ...151..547T}.
A wide range of codes are available to choose from (e.g. {\sc ProSpect}, \citealt{2020MNRAS.495..905R};  {\sc beagle}, \citealt{2016MNRAS.462.1415C}; \bagpipes, \citealt{2018MNRAS.480.4379C}; {\sc cigale}, \citealt{2009A&A...507.1793N}, \citealt{2019A&A...622A.103B}, \citealt{2020MNRAS.491..740Y}; {\sc prospector}, \citealt{2017ApJ...837..170L}, \citealt{2021ApJS..254...22J}; {\sc magphys}, \citealt{2008MNRAS.388.1595D}; {\sc bayeSED}, \citealt{2012ApJ...749..123H, 2014ApJS..215....2H, 2019ApJS..240....3H}), each employing differing underlying strategies to combine stellar evolution, dust and nebular models, and estimate the best-fitting parameters.
SED codes compare observed galaxy emission with theoretical templates to estimate physical properties.
To do so, they rely on assumptions about which input models (e.g. stellar evolution and spectral libraries, dust attenuation and emission models) and prior parameter ranges (e.g. age, mass, metallicity, initial mass function) to implement.
Such assumptions imply the observed galaxy is fully represented by the inputs and prior parameter space, potentially introducing inaccuracies when the galaxy's natural evolution deviates from this framework.

One critical component of SED fitting is the stellar population synthesis (SPS) framework, which describes how the emission of a population of stars evolves over time.
An SPS framework has three main ingredients: evolutionary tracks describing how individual stellar properties change with time, a stellar initial mass function (IMF) expressing the distribution of masses of stars at formation, and a set of stellar spectral templates to link stellar properties to an emission spectrum.
Each component can vary in formulation and can be hard to constrain from surveys.
The IMF can be derived through a model comparison to data, so most SPS frameworks have this as a tunable parameter.
However, the same is not true for stellar spectral library or evolutionary tracks, which require detailed modelling of individual stars to obtain rather than having imprints in global population properties.
Due to this, evolutionary tracks and stellar spectral libraries are generally hardcoded into SPS models and forgotten about when determining uncertainties.

With a wide grid of SPS component choices presented in literature and currently no constraints or consensus on which set of assumptions to apply, there is a large uncertainty associated with SPS frameworks that is overlooked too often.
Thus, we present an analysis to highlight the uncertainties introduced due to the hardcoded, highly uncertain parameter of the stellar spectral library by testing multiple libraries in one SPS framework and their impact on derived galaxy properties.
Alongside this, we also investigate the uncertainties which arises from two tunable, "known unknown" parameters: the IMF and metallicity assumptions.
This is done in the context of the well-established and extensively tested stellar evolution and population synthesis Binary Population and Spectral Synthesis framework \citep[BPASS,][]{2017PASA...34...58E, 2018MNRAS.479...75S} using both spectra and photometric simulated data.

The functional form of the IMF has been the subject of intense study for decades (e.g. \citealt{1955ApJ...121..161S, 1986FCPh...11....1S, 1998ApJ...504..835S, 2001MNRAS.322..231K, 2003PASP..115..763C, 2007ApJ...671..767T}; see \citealt{2018PASA...35...39H} for a review of measuring the IMF), yet it remains a major ``known unknown.''
While most local measurements support a near-universal IMF \citep[e.g.][]{2002Sci...295...82K, 2003PASP..115..763C, 2011ApJ...727...64K, 2011ASPC..448..361B, 2013pss5.book..115K}, other studies suggest it may vary with environment and redshift \citep[e.g.][]{2010ARA&A..48..339B, 2012Natur.484..485C, 2012MNRAS.422.2246M, 2020ARA&A..58..577S, 2013ApJ...771...29G, 2014ApJ...796...71Z, 2018ApJ...857...46I, 2023MNRAS.519.4753T, 2023arXiv231102051C}.
Studies have shown that fitting observational data with an IMF prescription inconsistent with the true underlying distribution can significantly bias the inferred parameters, such as age, mass, and metallicity \citep[e.g.][]{2023arXiv231018464W, 2024A&A...686A.138C, 2024ApJ...963...74W}.

Similarly, a wide range of stellar spectral libraries are available for use in SPS frameworks \citep[e.g.][]{2000MNRAS.315..679C, 2002A&A...381..524W, 2011A&A...532A..95F, 2014MNRAS.440.1027C, 2014ApJ...780...33C, 2016ApJ...823..102C, 2018A&A...618A..25A, 2021MNRAS.504.2286K}.
Each varies in terms of the input physics used to model stellar atmospheres and, as a result, can produce significantly different spectra \citep[e.g.][]{2023MNRAS.521.4995B}.
These differences stem from limited observational constraints and incomplete or often inaccurate physics (i.e. atomic and molecular line lists), meaning high confidence spectra are not always available.
Consequently, spectral libraries often vary in their wavelength and spectral type coverages, and may disagree on the spectrum associated with a star of given luminosity, temperature, or gravity.
This disagreement propagates into SED fitting and has been shown to influence the inferred parameter space \citep[e.g.][]{2009ApJ...699..486C, 2020MNRAS.491.2025C, 2023MNRAS.521.4995B}.

In addition to uncertainties inherent to the construction of SPS models, prior assumptions made during SED fitting also introduce significant sources of systematic error, such as the choice of metallicity.
These, similarly to IMF, are "known unknowns" since global galaxy parameters can be used to constrain them, and their associated uncertainties are considered during fitting, unlike those arising due to stellar spectral library.
SPS templates are constructed at a range of metallicities to reflect the chemical enrichment history of the Universe.
The first stars formed in a metal-free environment, and through nucleosynthesis, progressively enriched the interstellar medium \citep[e.g.][]{2015ApJ...800...20G, 2022MNRAS.514.1315B}.
However, galaxy metallicities at any given cosmic epoch exhibit substantial scatter, and inferring metallicity from observations is challenging due to degeneracies with other parameters, such as stellar age and dust attenuation \citep[e.g.][]{1994ApJS...95..107W}.
As a result, metallicity is often treated as a nuisance parameter during SED fitting and is frequently fixed to remain constant over a galaxy’s history \citep[e.g.][]{2010A&A...523A..13P, 2015MNRAS.447..786P, 2023A&A...669A..11P}.
While this is a common practice, it is an oversimplification that can introduce significant systematic uncertainties.
Accurate metallicity estimates are essential for reliably deriving galaxy properties, and neglecting this can skew interpretations of fundamental parameters \citep[e.g.][]{2024arXiv241017698B}.

If such variations affect the inferred properties of individual galaxies, they may also influence any global quantity derived from ensembles of galaxies.
One key diagnostic of galaxy evolution is the cosmic star formation rate density (CSFRD), alongside the evolution of the cosmic stellar mass density.
The CSFRD describes the total star formation rate (SFR) across all galaxies per unit co-moving volume as a function of time.
The integral of this over time (after accounting for the death of massive stars) should match the in-situ volume-averaged stellar mass density at each redshift.
Extensive investigations have been conducted to derive both the CSFRD \citep[e.g.][]{1996ApJ...460L...1L, 1996MNRAS.283.1388M, 1998ApJ...498..106M, 1997AJ....113....1S, 2005ApJ...632..169L, 2005ApJ...619L..47S, 2006ApJ...651..142H, 2011MNRAS.413.2570R, 2012A&A...539A..31C, 2013MNRAS.428.1128S, 2016MNRAS.461..458D, 2016MNRAS.461.1100R, 2019A&A...624A..98W, 2024MNRAS.527.5525K} and stellar mass assembly history \citep[e.g.][]{2003ApJ...587...25D, 2005ApJ...619L.131D, 2007A&A...474..443P, 2008ApJ...675..234P, 2011MNRAS.413..162C, 2013A&A...556A..55I, 2013ApJ...777...18M, 2015ApJ...801...97S, 2017A&A...605A..70D, 2018MNRAS.475.2891D, 2025arXiv250303431D} of galaxies.
However, it has been shown that the two are inconsistent \citep[e.g.][]{2008MNRAS.385..687W, 2014ARA&A..52..415M, 2016ApJ...820..114Y, 2018PASA...35...39H, 2019ApJ...877..140L, 2019MNRAS.490.5359W}.
Possible explanations include issues with dust corrections, inappropriate or variable initial mass function (IMF), incorrect inferred metallicities, or requiring the application of alternate stellar population synthesis models \citep{2016ApJ...820..114Y, 2019MNRAS.490.5359W}.

Numerical simulations (e.g. Millennium, \citealt{2005Natur.435..629S}, \citealt{2009MNRAS.398.1150B}, \citealt{2012MNRAS.426.2046A}; Illustris, \citealt{2014MNRAS.444.1518V}, \citealt{2018MNRAS.475..676S}; EAGLE, \citealt{2015MNRAS.450.1937C}, \citealt{2015MNRAS.446..521S}) offer a controlled means of investigating galaxy formation and evolution, allowing us to test our understanding of global quantities such as the CSFRD.
They produce mock galaxy populations with known masses, metallicities, and star formation histories.
These can be translated into synthetic observed emission spectra with the application of a stellar population synthesis template, assuming consistency between the simulation's input physics and those used in the SPS model.
If a single SPS framework is used to generate the mock observed spectrum, SED fitting then enables an investigation of how adopting different SPS models, each with their own assumptions, affects the recovery of physical parameters.
This approach allows us to quantify systematic modelling uncertainties in a controlled setting.

In this work, we employ a suite of simulated galaxies as a controlled environment for testing the impact of differing SPS assumptions.
Mock spectra are generated using a single SPS framework, and then fit using alternative templates that differ in their choice of stellar library, IMF, and metallicity prescription.
We examine how these choices shift the recovered parameter space and quantify their impact on the uncertainty budget.
Section~\ref{sec:tsc_mods} introduces the SPS variants compared in this study within the BPASS framework.
Section~\ref{sec:tsc_meth} describes the methodology to convert simulated particle data into synthetic integrated light observed galaxy emission spectra, as well as the SED spectral and photometric fitting technique.
The effects of varying the spectral library, IMF, and metallicity prescriptions are presented in Sections~\ref{sec:spec_lib_impact}-\ref{sec:met_impact}, respectively.
Section~\ref{sec:mah_impact} explores the consequences for the inferred mass assembly history of galaxies, and
Section~\ref{sec:tsc_consequences} discusses implications for survey-based studies of galaxy evolution.
We present a brief summary of our main findings in Section~\ref{sec:tsc_concs}.

\section{Models} \label{sec:tsc_mods}
Our analysis focuses on highlighting the uncertainties arising from the hardcoded parameter of the stellar spectral library, to identify how extreme the uncertainties from this are relative to observational and other fitting uncertainties.
We also explore the initial mass function and metallicity prescription assumptions as these "known unknown" parameters are generally the least well-constrained.
We use the Binary Population And Spectral Synthesis framework \citep[BPASS,][]{2017PASA...34...58E} as the basis for each stellar population.
This is a set of SPS models which incorporate binary evolution in addition to single star evolutionary pathways, tracing the evolution of simple stellar populations from an age of 1\,Myr up to 100\,Gyr.
The binary models are full detailed stellar evolution models, in which the interior structure of the star is calculated, and which allow for mass-loss or gain through binary interactions.
The same metallicity-dependent stellar evolution models and binary parameter assumptions underlie all synthetic populations constructed in this work.
A investigation into the uncertainty arising from using single-only evolutionary tracks versus binary-incorporated tracks on derived galaxy parameters is presented in Appendix~\ref{sec:evo_tracks}.

The BPASS framework has generated separate versions that account for variations in these assumptions.
The v2.2.1 models \citep{2018MNRAS.479...75S} are used as the standard comparison set and incorporate different IMF forms (see Section~\ref{sec:tsc_imf_vars}).
The v2.3.1 models \citep{2022MNRAS.512.5329B, 2023MNRAS.521.4995B} include SPS models built using various stellar spectral templates (see Section~\ref{sec:tsc_spec_libs}).
When testing IMF and spectral template variations, the metallicity for each model is fixed.
To explore the impact of this fixed assumption, we conduct two separate metallicity assumption tests, with details provided in Section~\ref{sec:tsc_bagpipes}.
This work neglects nebular emission.
Unless otherwise stated, this work and the BPASS models assume that the Sun has a metallicity value of $Z_\odot=0.020$

\subsection{IMF Variations} \label{sec:tsc_imf_vars}
BPASS v2.2.1 defines its default IMF following \citet{1993MNRAS.262..545K} using a broken power-law function.
This has a lower IMF slope, $\alpha_1$, applied between 0.1\,M$_\odot$ and $M_1$ and an upper slope, $\alpha_2$, applied between $M_1$ and $M_{\mathrm{max}}$, i.e.
\begin{equation}
\begin{split}
    N(M<M_{\mathrm{max}}) \propto & \int_{0.1}^{M_1} \left( \frac{M}{\mathrm{M}_\odot} \right)^{\alpha_1} dM \\
    & + M_1^{\alpha_1-\alpha_2} \int_{M_1}^{M_{\mathrm{max}}} \left( \frac{M}{\mathrm{M}_\odot} \right)^{\alpha_2} dM.
\end{split}
\end{equation}
The default BPASS model, used as the comparison standard in this work, sets $\alpha_1=-1.30$, $\alpha_2=-2.35$, $M_1=0.5$\,M$_\odot$ and $M_{\mathrm{max}}=300$\,M$_\odot$.
BPASS v2.2.1 also contains other stellar population models generated with different IMF parameters.
This work compares the results of variations in the upper slope: one steeper ($\alpha_2=-2.70$), one with a shallower upper slope ($\alpha_2=-2.00$), and one which has the same slope for both power laws (i.e. continuous) and a lower maximum stellar mass ($\alpha_1=\alpha_2=-2.35$, $M_{\mathrm{max}}=100$\,M$_\odot$) following the prescription of \citet{1955ApJ...121..161S}.

In addition, we consider one IMF which, rather than using a broken power law, has an exponential cut-off at masses below 1\,M$_\odot$, following the prescription of \citet[][hereafter C03]{2003PASP..115..763C}.
This has a slightly shallower upper slope of $\alpha_2=-2.3$ compared to the BPASS default and an upper mass limit of $M_{\mathrm{max}}=300$\,M$_\odot$.
These models, along with labels used throughout this work, are presented in Table~\ref{tab:imf_vars}.
The IMF slopes and turnovers are representative of the range of behaviour seen in local ($z=0$) stellar populations \citep[see][for an informative compilation]{2024ARA&A..62...63H}.
There is also some evidence that intense starbursts more reliably populate the upper end of the stellar initial mass function than ongoing low level star formation \citep[see e.g.][]{2024arXiv241007311K}. Hence galaxies which formed their stars in an intense burst at high redshift, although quiescent when observed at z=0, may encode the fossil of more top-heavy IMFs (shallower slopes) in their surviving stellar population. Conversely, star-forming galaxies in the local Universe, which typically have low star formation rates, may be relatively top-light.

\begin{table}
    \begin{center}
    \caption[Initial mass function prescriptions adopted in this paper.]{IMF model variations tested in this work and the label used for each.
    `Exp cut-off' indicates an exponential cut-off in the mass distribution following the prescription of \citet{2003PASP..115..763C}.}
    \label{tab:imf_vars}
    \begin{tabular}{lcccc}
        \hline
        Model Label & $\alpha_1$ & $\alpha_2$ & $M_1$ / M$_\odot$ & $M_{\mathrm{max}}$ / M$_\odot$ \\
        \hline \hline
        Default & -1.30 & -2.35 & 0.5 & 300 \\
        Shallow & -1.30 & -2.00 & 0.5 & 300 \\
        Steep & -1.30 & -2.70 & 0.5 & 300 \\
        Continuous & -2.35 & -2.35 & 0.5 & 100 \\
        C03 & exp cut-off & -2.3 & 1.0 & 300\\
        \hline
    \end{tabular}
    \end{center}
\end{table}

\subsection{Stellar Spectral Libraries} \label{sec:tsc_spec_libs}
The BPASS stellar spectral grid is formed from multiple libraries depending on the type of star.
For v2.2.1, the main-sequence and giant branch stars are drawn from the \citet{1970SAOSR.309.....K, 1993sssp.book.....K} models of \citet{2014ApJ...780...33C}.
This is supplemented with Wolf-Rayet models \citep{2003A&A...410..993H, 2015A&A...577A..13S} with an excess He{\sc ii} flux  due to under-representation \citep[as suggested by][]{2008MNRAS.385..769B}, {\sc wm-basic} \citep{1998ASPC..131..258P} O star models based upon the grid of surface parameters determined by \citet{2002MNRAS.337.1309S}, and a synthetic spectral grid of white dwarf atmospheres from \citet{2017ApJS..231....1L}.

Four alternative stellar spectral libraries implemented in BPASS v2.3.1 are used in this work, namely the \citet[][hereafter AP]{2018A&A...618A..25A}, BaSeL \citep{1998A&AS..130...65L, 2002A&A...381..524W}, CKC \citep{2014ApJ...780...33C} and C3K \citep{2016ApJ...823..102C} libraries, each replacing the main sequence-giant branch grid.
All use the BPASS framework with a default BPASS IMF between $0.1-300$\,M$_\odot$, the same underlying stellar evolution models and population synthesis but apply different models during spectral synthesis. Hereafter we refer to the grid of SPS models generated using a given stellar spectral template library by the name of that library.
Stellar evolution models are matched to the nearest spectral template available based on the library's temperature, surface gravity and composition grid.
This is achieved by interpolating between spectra in the library which bracket the evolution model.
If the model is outside the spectral template's parameter space, then the nearest template is assigned rather than extrapolating.
All resolutions are degraded to a fixed $1$\,\AA\ resolution.
\citet{2023MNRAS.521.4995B} define this as the BPASS v2.3.1 framework.
The important characteristics behind each library are given in Table~1 of \citet{2023MNRAS.521.4995B} and summarised below.

\subsubsection{AP - \citep{2018A&A...618A..25A}}
The \citet{2018A&A...618A..25A} spectral library has a wavelength coverage of 2000\,\AA\ to 2.5\,$\mu$m at a resolution of $R\simeq30\,000$ with a broad composition parameter range, including spectra covering temperatures of $3500-30\,000$\,K.
The BPASS framework uses non-local thermal equilibrium (non-LTE) OB stellar spectra calculated using WM-BASIC \citep{1998ASPC..131..258P} for main sequence stars hotter than 25\,000\,K, meaning the hottest spectra from the AP library are not included.
\citet{2018A&A...618A..25A} computed their spectra templates using the {\sc atlas9} \citep{1992IAUS..149..225K} model atmospheres of \citet{2012AJ....144..120M}.

\subsubsection{BaSeL - \citep{2002A&A...381..524W}}
The BaSeL library has been widely used in population synthesis codes and covers a broad region of parameter space, including stellar atmospheric temperatures of $2000-50\,000$\,K.
BPASS used the BaSeL main sequence spectral libraries up to 25\,000\,K before switching to the non-LTE OB grid in all versions up to and including v2.1 \citep{2017PASA...34...58E}.
The library has a wavelength coverage of $\sim1785-100\,000$\,\AA, with a spectral resolution of $R\simeq110$ (in the optical) which is rather poor compared to more recent models.

\subsubsection{CKC - \citep{2014ApJ...780...33C}}
The CKC library replaced the BaSeL model for the main sequence and giant branch spectral library from BPASS v2.2 onwards \citep{2018MNRAS.479...75S} and acts as the best comparison between the control v2.2.1 and the v2.3.1 frameworks.
Stellar atmospheres and spectra were constructed using the {\sc atlas12} \citep{2005MSAIS...8...25C} and {\sc synthe} \citep{1993sssp.book.....K} codes, respectively.
These models cover a similar temperature and gravity parameter space to the BaSeL grid, but over a wavelength range that goes bluer ($100-100\,000$\,\AA) with a higher resolution of $R=3000$.
While these should match the control v2.2.1 framework, small differences in bolometric luminosity normalisation and smoothing mean both are tested for complete comparison.

\subsubsection{C3K - \citep{2016ApJ...823..102C}}
The C3K library is the successor to CKC, utilising {\sc atlas12} and {\sc synthe}, and were introduced to BPASS in the v2.3 framework \citep{2022MNRAS.512.5329B}.
As with the AP library, the hottest stars in the C3K library are replaced with non-LTE spectra reflecting strong winds.
The updated C3K grid has a limited wavelength coverage compared to the CKC library ($100-20\,000$\,\AA) but covers this at a higher resolution of $R=10\,000$ (in the optical).

\subsection{BPASS Model Spectra Comparison} \label{sec:tsc_bpass_mod_comp}
\begin{figure}
    \centering
    \includegraphics[width=\columnwidth]{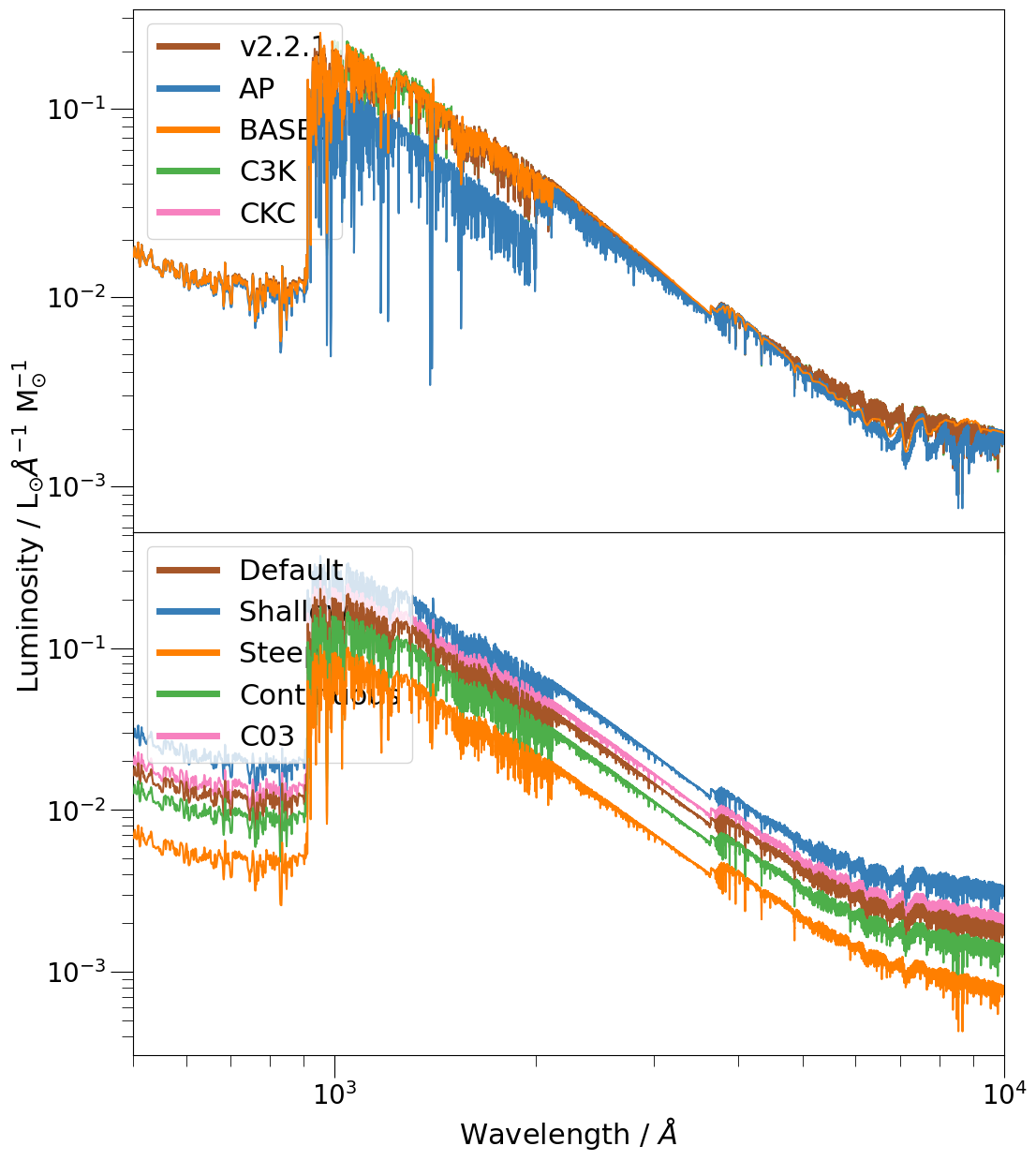} 
    \caption[BPASS template models generated using different atmospheric templates and IMF prescriptions for a 10\,Myr old stellar population.]{Spectra generated from the BPASS framework using different stellar spectral libraries (top panel) and IMF prescriptions (bottom panel) for a 10\,Myr old stellar population.
    The top panel includes v2.2.1 in brown, AP in blue, BaSeL in orange, C3K in green and CKC in pink.
    The bottom panel shows, in order from top to bottom, the shallow IMF in blue, C03 prescription in pink, default in brown, continuous in green, and steep IMF in orange.}
    \label{fig:spec_comp_7}
\end{figure}

\begin{figure}
    \centering
    \includegraphics[width=\columnwidth]{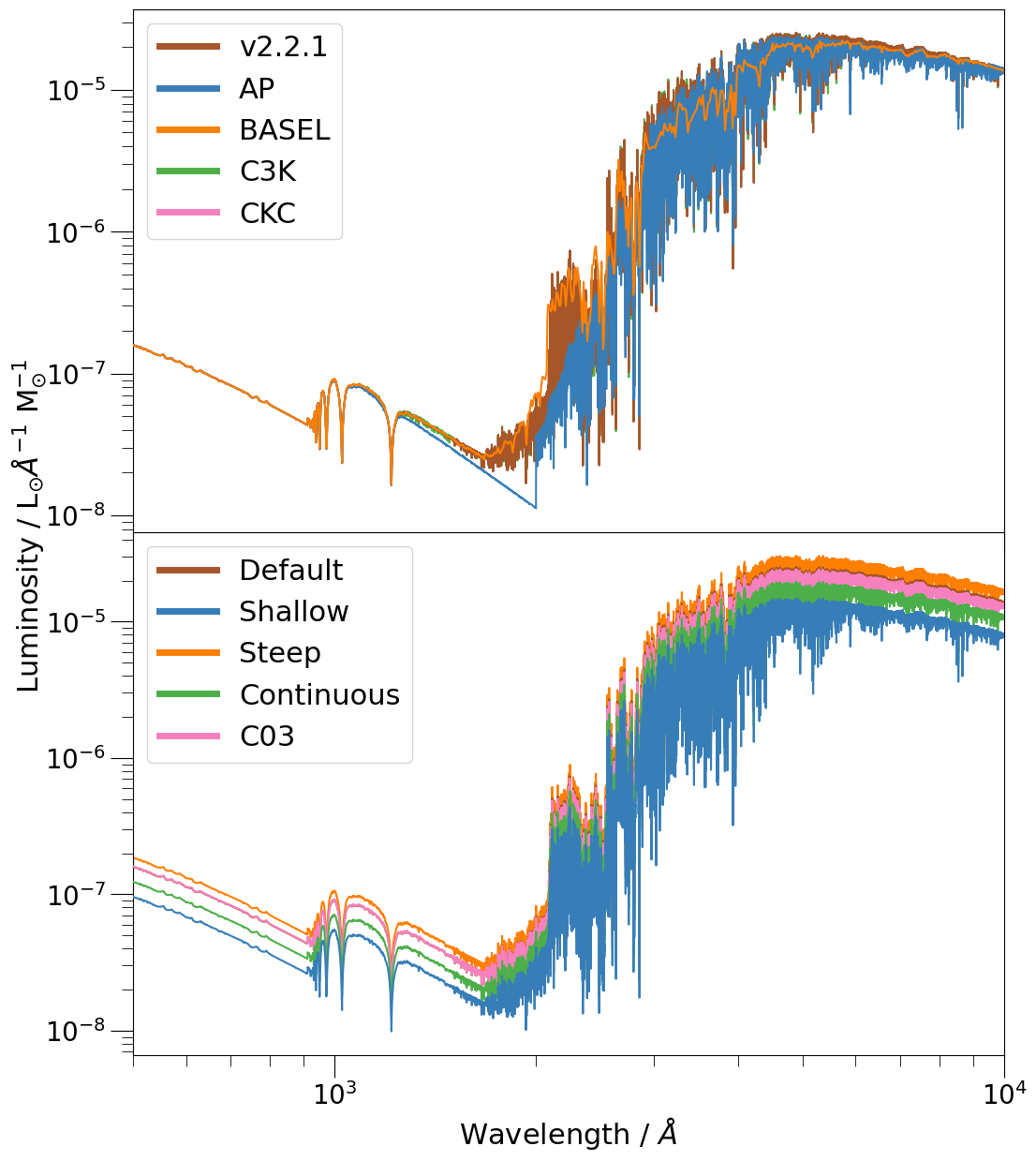} 
    \caption[]{Same as Fig.~\ref{fig:spec_comp_7} for a 10\,Gyr old stellar population.
    The top panel includes the stellar spectral libraries of v2.2.1 in brown, AP in blue, BaSeL in orange, C3K in green and CKC in pink.
    The bottom panel shows, in order from top to bottom, the steep IMF in orange, default in brown, C03 prescription in pink, continuous in green, and shallow IMF in blue.
    A UV upturn is observed, which arises naturally in binary models due to compact, accreting hot sub-dwarfs and white dwarfs.}
    \label{fig:spec_comp_10}
\end{figure}

The integrated optical spectra of simple (i.e. coeval) stellar populations generated when applying the different stellar template libraries (top panel) and IMF prescriptions (bottom panel) are shown in Figs~\ref{fig:spec_comp_7} and \ref{fig:spec_comp_10} for a 10\,Myr and 10\,Gyr stellar population, respectively.
Little difference is observed between v2.2.1, C3K and CKC spectra since these are near-identical templates.
More substantial differences are observed when comparing to the AP and BaSeL spectra.
BaSeL templates have lower resolution and fail to reproduce the narrow line features resolved by the other models.
The AP spectra predict lower fluxes for both stellar population ages presented.
This will affect fitting to galaxy spectra generated using the v2.2.1 framework as the imbalance will need to be corrected by altering extinction or introducing a young stellar population.

The IMF prescriptions considered result in spectra with similar slopes and features since they utilise the same spectral library, but flux offsets arise due to mass being influenced by the relative number of faint stars.
In the 10\,Myr stellar population, short-lived massive and blue stars dominate the flux, while low mass stars dominate the mass.
Shallower upper-mass slopes produce more high-mass, luminous stars and so generate brighter spectra for a given mass of star formation.
Fig.~\ref{fig:spec_comp_7} highlights this with the the shallow ($\alpha=-2.00$) and steep ($\alpha=-2.70$) prescription generating the most and least luminous spectra, respectively.
The C03 IMF has a slightly shallower upper slope than the default value causing it to be slightly more luminous.
On the other hand, the continuous prescription has the same upper slope but contains a higher fraction of low-mass stars and ignores stars of masses $100-300$\,M$_\odot$, resulting in a lower flux normalisation for a given total mass of star formation.

For the 10\,Gyr stellar population, the reverse is true: steeper upper-mass prescriptions now generate the most luminous spectra for a given initial mass of star formation.
Steeper slopes result in a higher fraction of low-mass stars per unit mass, while shallower slopes allocate more mass to high-mass stars.
Old, 10\,Gyr stellar populations have lost all high-mass, short-lived stars and so have spectra dominated by the low-mass stellar light.
With the continuous prescription, the steeper low-mass slope relative to the default framework generates a lower flux normalisation.

\section{Methodology} \label{sec:tsc_meth}
To test the impact of spectral template variations, we need to construct realistic galaxies in which the star formation history and metal enrichment history is known to a high degree of confidence.
A sample constructed using simulated data,  allows for the generation of mock observations from SPS models with known physics and assumptions.
This enables us to investigate variations in inferred galaxy properties when using models with inaccurate assumptions.
We draw simulated data  from the Evolution and Assembly of GaLaxies and their Environments project \citep[EAGLE,][described in more detail in section~\ref{sec:mgs}]{2015MNRAS.450.1937C}.

\begin{figure*}
    \centering
    \includegraphics[width=\textwidth]{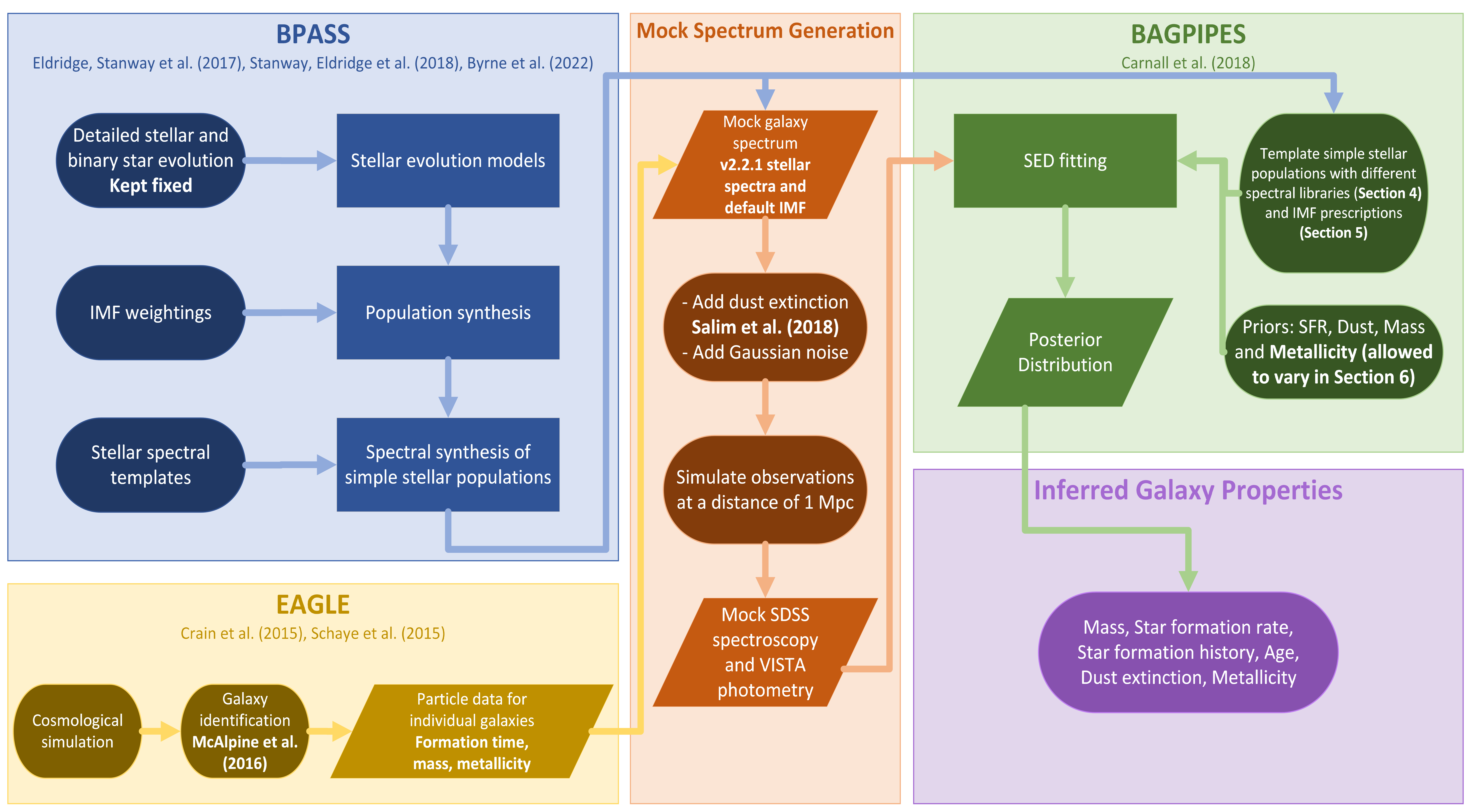}
    \caption[]{Flowchart outlining the methodology used in our work.
    The blue box (upper left) describes the BPASS framework and the steps taken to generate simple stellar population models (Section~\ref{sec:tsc_mods}).
    Different IMF and stellar spectral libraries are applied during the population and spectral synthesis steps, respectively, to generate models with varying underlying assumptions.
    The yellow box (lower left) represents our simulated galaxy sample from the EAGLE project and the data acquired to generate mock observations (Section~\ref{sec:mgs}).
    This data is combined with the default BPASS v2.2.1 framework to construct mock galaxy observations in the orange box (center, Section~\ref{sec:spec_gen}).
    Comparison between the mock galaxy observations and all BPASS templates is performed using the \bagpipes\ SED fitting software, highlighted in the green box (upper right, Section~\ref{sec:tsc_bagpipes}).
    The process from from simple to composite stellar populations is outlined with the assumptions being varied highlighted.
    This generates the inferred galaxy properties in the purple box (lower right).
    }
    \label{fig:flowchart}
\end{figure*}

A flowchart outlining the method is shown in Fig.~\ref{fig:flowchart}.
The blue box highlights the BPASS framework, introduced in Section~\ref{sec:tsc_mods}, and the points where different stellar libraries and IMF prescriptions are incorporated.
Yellow is the simulated data drawn from EAGLE, which is combined with the default v2.2.1 BPASS framework in the orange box to generate our mock observations (Section~\ref{sec:spec_gen}).
Comparison between mock observations and BPASS templates is performed in the green box using the Bayesian Analysis of Galaxies for Physical Inference and Parameter EStimation SED fitting software (\bagpipes, Section~\ref{sec:tsc_bagpipes}), resulting in the inferred galaxy properties shown in purple.

\subsection{Mock Galaxy Sample} \label{sec:mgs}
A sample of simulated $z=0$ galaxies are acquired from the Evolution and Assembly of GaLaxies and their Environments project \citep[EAGLE,][]{2015MNRAS.450.1937C, 2015MNRAS.446..521S}.
EAGLE is a set of state-of-the-art, large-scale hydrodynamical simulations of a $\Lambda$-cold dark matter (CDM) universe examining the formation of galaxies and their co-evolution with their gaseous environments.
They follow the evolution of galaxies from $z=127$ to the present day, calibrated to match the $z=0.1$ galaxy stellar mass function, the $z\sim0$ star-forming galaxy stellar mass-size relation, and the stellar mass to super massive black hole mass relation.
The simulations utilised an extensively modified version of the $N$-Body smoothed particle hydrodynamics code {\sc gadget-3} \citep[][]{2005MNRAS.364.1105S}, including subgrid treatments of baryonic physics assuming a \citet{2003PASP..115..763C} IMF.
The subgrid physics includes prescriptions for radiative cooling, star formation, stellar mass loss, metal enrichment, stellar and AGN feedback, and supermassive black hole accretion and mergers.
EAGLE adopts a Planck cosmology where $\Omega_\Lambda=0.693$, $\Omega_m=0.307$, $\Omega_b=0.04825$ and $H_0=67.77$\,km\,s$^{-1}$Mpc$^{-1}$ \citep{2014A&A...571A...1P}.

The EAGLE suite has a range of simulations with different resolutions, implemented subgrid models, and cubic volumes ranging from 25 to 100 comoving Mpc.
For this work, we use the EAGLE simulation L0025N0376.
This has a box side length of 25 comoving Mpc and contains 376$^3$ particles of both dark matter and baryons.
The dark matter has an initial particle mass of $9.70\times10^6$\,M$_\odot$ and the initial gas (baryon) particle mass is $1.81\times10^6$\,M$_\odot$.
The BPASS models are generated to represent the average stellar emission from a $10^6$\,M$_\odot$ population, which matches closely to the mass resolution of the box.
Going to a higher mass resolution would make stochastic sampling of the IMF a consideration, where the stellar emission from a population may differ significantly from the fiducial BPASS models \citep[see][]{2023MNRAS.522.4430S}.

Each baryonic matter particle in the simulation has a unique identifier which allows it to be tracked throughout the simulation at different times.
To generate galaxy catalogs, the EAGLE team used the {\sc subfind} algorithm \citep{2001MNRAS.328..726S, 2009MNRAS.399..497D} which identified gravitationally bound subhaloes as individual galaxies.
The catalogues generated through this procedure form the public data released which were queried through an SQL database.
We select galaxies from the snapshot corresponding to $z=0$ with masses $M>10^9$\,M$_\odot$.
This will select predominantly old galaxies, permitting investigation into the star formation and mass assembly histories of galaxies from their formation to present day.
The lower mass limit reduces variation in metallicity due to the mass-metallicity relation, and chooses a group of typical old, massive galaxies that are a mixture of quiescent and star-forming.
The EAGLE team assigned a random number between 0 and 1 to each galaxy.
We select the 18 galaxies with the lowest random numbers as test cases.
The IDs and properties of the selected galaxies are listed in Table~\ref{tab:galaxy_ids}.

\begin{table}
    \begin{center}
    \caption[]{EAGLE galaxy IDs used in this analysis and their key properties, arranged in order from lowest random number.
    The age is weighted by the current living stellar mass in each particle.
    The mean metallicity is also stellar mass weighted.}
    \label{tab:galaxy_ids}
    \begin{tabular}{l cccc}
        \hline
        \multicolumn{1}{c}{Galaxy} & Stellar Mass / & Mass Weighted & SFR / & Mean Stellar \\
        \multicolumn{1}{c}{ID} & $10^{10}$\,M$_\odot$ & Age / Gyr & M$_\odot$ yr$^{-1}$ & Metallicity, Z \\
        \hline \hline
        233212 & 0.13 & 7.22 & 0.029 & 0.0147 \\
        345925 & 15.4 & 8.61 & 1.907 & 0.0166 \\
        191315 & 1.04 & 8.40 & 0.240 & 0.0205 \\
        826275 & 0.24 & 6.72 & 0.220 & 0.0198 \\
        225579 & 0.11 & 6.93 & 0.000 & 0.0144 \\
        221438 & 0.31 & 8.45 & 0.072 & 0.0172 \\
        004716 & 0.21 & 5.52 & 0.391 & 0.0158 \\
        179177 & 1.43 & 5.98 & 1.021 & 0.0180 \\
        197376 & 0.42 & 8.61 & 0.014 & 0.0137 \\
        223121 & 0.15 & 8.19 & 0.091 & 0.0176 \\
        832048 & 0.41 & 5.88 & 0.394 & 0.0207 \\
        148609 & 2.18 & 7.49 & 1.419 & 0.0185 \\
        232726 & 0.13 & 7.75 & 0.086 & 0.0170 \\
        141928 & 3.50 & 6.11 & 5.133 & 0.0238 \\
        219248 & 0.13 & 8.03 & 0.133 & 0.0153 \\
        258945 & 6.02 & 9.58 & 2.070 & 0.0196 \\
        009478 & 1.24 & 6.04 & 0.757 & 0.0200 \\
        193121 & 0.98 & 6.92 & 0.348 & 0.0179 \\
        \hline
    \end{tabular}
    \end{center}
\end{table}

\subsection{Spectra Generation} \label{sec:spec_gen}
The EAGLE team provide both halo catalogue and individual particle data in their data release (\citealt{2016A&C....15...72M}; also see \citealt{2017arXiv170609899T}).
For each galaxy chosen from the halo catalogue, the particles assigned to that galaxy are identified.
Each particle contains information on its current metallicity, birth mass and its formation time.
We use this to assign the closest matching SPS template within the BPASS grid, scaling the luminosity by the particle's mass.
Summing the individual SPS templates for all particles within the galaxy generates a mock emitted spectrum for that galaxy.
Each galaxy spectrum is generated using the v2.2.1 default BPASS framework, generating 18 spectra in total.
We therefore assume that a galaxy's emission spectrum is well approximated by the BPASS v2.2.1 models.
For validation of the spectra generation code, see Appendix~\ref{sec:code_cal}.

Each spectrum is converted into mock observations designed to replicate real observational data.
We simulated observations using Sloan Digital Sky Survey \citep[SDSS,][]{2000AJ....120.1579Y, 2006AJ....131.2332G, 2009ApJS..182..543A} spectra and photometry, and Visible and Infrared Survey Telescope for Astronomy \citep[VISTA,][]{2015A&A...575A..25S} photometry.
SDSS uses a wide-field ground-based 2.5\,m telescope with two instruments: an imager \citep{1998AJ....116.3040G} utilising the five filters \textit{ugriz} and a spectrograph with wavelength coverage from 3800 to 9200\,\AA\ at a resolution of $\lambda/\Delta\lambda\simeq2000$.
VISTA is a survey telescope with a camera operating at wavelengths $0.8-2.3$\,$\mu$m.
We simulate VISTA photometry using the Y, J and H bands.
We do not include the Ks band due to limited wavelength coverage of select spectral libraries considered.

To replicate observations, attenuation of starlight by dust must be incorporated.
Attenuation is applied to each spectrum assuming a \citet{2018ApJ...859...11S} attenuation curve.
\citet{2018ApJ...859...11S} studied the dust attenuation curves of 230,000 galaxies in the local Universe using GALEX, SDSS and WISE photometry and the {\sc cigale} SED-fitting code \citep{2009A&A...507.1793N, 2019A&A...622A.103B}. 
Their sample included a wide range of galaxy types, ranging from quiescent to intensely star-forming systems.
They utilise this to generate dust attenuation curves for various galaxy types and as a function of mass.
In this work, we take the two overall attenuation curves for star-forming galaxies and quiescent galaxies.
Previous work suggests that galaxies can be distinguished between quiescent and star-forming by a rest-frame two- or three-band colour selection \citep[e.g.][]{2001AJ....122.1861S, 2007ApJ...655...51W, 2016ApJ...830...51S, 2021ApJ...923...46L}, or by their specific star formation rate (sSFR) with agreement between the two \citep{2013A&A...556A..55I}.
We follow the work of \citet{2010ApJ...709..644I} and \citet{2011MNRAS.417..900D} to apply the quiescent attenuation law to any galaxy with $\log(\mathrm{sSFR} / \mathrm{yr}^{-1})<-11$ and a star-forming law to the rest.
We assume a uniform total extinction of $\mathrm{E(B-V)}=0.3$.
The impact of including dust attenuation in our methodology is discussed in Appendix~\ref{sec:impact_of_dust}.

All galaxies are redshifted to $z=0.00023$ to be placed at an effective distance of 1\,Mpc.
We assume a standard $\Lambda$-CDM cosmological model in which $\Omega_\Lambda=0.7$, $\Omega_{\mathrm{M}}=0.3$, and $H_0=70$\,km s$^{-1}$ Mpc$^{-1}$, which is also used by \bagpipes\ SED fitting software.
Random noise is applied to all data measurements following a normal Gaussian distribution.
This is done by generating a random noise element at each wavelength in the spectrum and for each photometry data point.
Median errors are set as 3\% of the original flux value.

\begin{figure}
    \centering
    \includegraphics[width=\columnwidth]{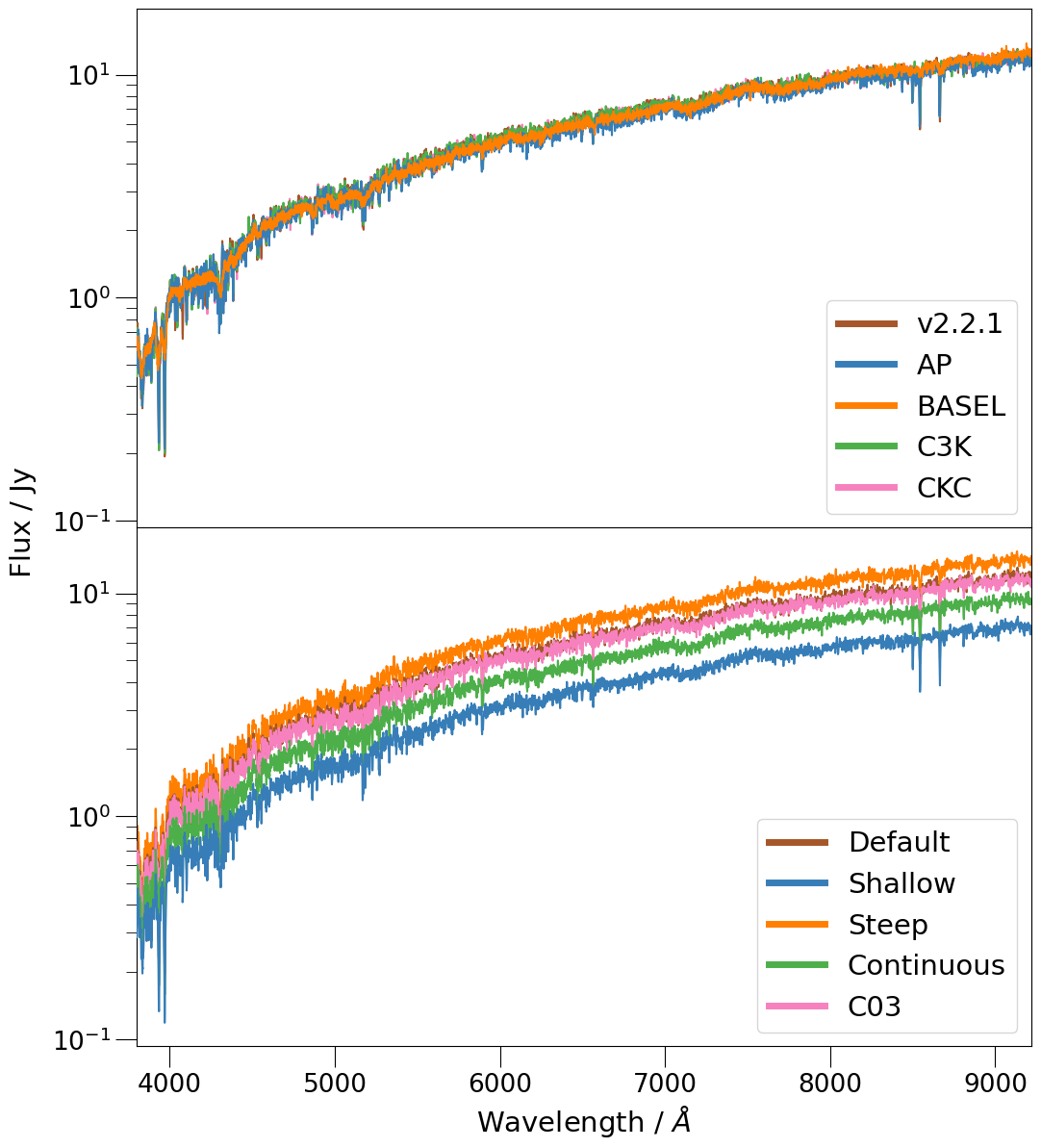} 
    \caption[]{Illustration of mock SDSS spectra produced using simulated galaxy ID 197376 when applying different stellar spectral libraries (top panel) and IMF prescriptions (bottom panel).
    Top panel shows the spectral libraries of v2.2.1 in brown, AP in blue, BaSeL in orange, C3K in green and CKC in pink.
    Bottom panel shows, in order from top to bottom, the steep prescription in orange, default BPASS IMF in brown, C03 in pink, continuous in green, and the shallow IMF in blue.}
    \label{fig:gal_spec_comp}
\end{figure}

As an exercise to illustrate the differences between tested model assumptions, we generate a template spectrum using the star-formation and metallicity history of simulated galaxy ID 197376 applying all spectral library and IMF prescriptions.
The simulated SDSS emission spectra are plotted in Fig.~\ref{fig:gal_spec_comp}.
The difference between the spectral library templates are minimal in this case.
The BaSeL template has the lowest resolution, which means it does not produce any narrow line features and will cause issues when comparing to other spectral templates which produce line features.
The difference between the IMF prescription spectra is more significant.
The steep IMF is the most luminous simulated observation while shallow IMF is least luminous, since this is a galaxy in which the old stellar population dominates stellar emission.

\subsection{SED Fitting with \bagpipes} \label{sec:tsc_bagpipes}
The 18 mock galaxy observations, were fitted with templates constructed using all other BPASS variants, using the Bayesian Analysis of Galaxies for Physical Inference and Parameter EStimation SED fitting software \citep[\bagpipes,][]{2018MNRAS.480.4379C}. In other words we are testing the scenario that the galaxy is truly described by BPASS v2.2.1 but, as observers in ignorance of this, we apply a different set of stellar populations in our fitting.
This incorporates a full spectral fit, using SDSS spectra from 3800 to 9200\,\AA\ and VISTA photometry in the Y, J and H bands between $0.8-2.3\,\mu$m.

The \bagpipes\ Bayesian spectral fitting code utilises a chosen stellar population synthesis model in combination with nebular emission libraries, a dust attenuation prescription, and a multi-component dust emission model, to fit and interpret the integrated light from a galaxy from the far-ultraviolet (UV) to the microwave regimes.
It returns posterior probability distributions for a range of galaxy properties including stellar mass, age, SFR, SFH, dust extinction and metallicity.

We use a modified version of the software to incorporate the \citet{2018ApJ...859...11S} dust attenuation law for star-forming and quiescent galaxies. Nebular attenuation and emission are neglected in both model construction and fitting.
BPASS SPS templates generated under each prior assumption are inserted into the modified \bagpipes\ code, and we fit all 18 galaxy spectra with each spectral library and IMF variant.
All fitting priors are identical in order to uniformly test the SPS model assumptions.
Each fit assumes that we, as observers, have acquired the simulated optical spectrum and Y, J and H band photometry, but have no further information.

To simulate the stochastic star formation histories of galaxies, a "non-parametric" SFH is implemented.
Non-parametric SFHs enable a broader range of star formation histories than parametric models, but come at the cost of increased computational expense.
The simplest approach defines a series of piecewise constant functions (i.e. step functions) that represent the mass formed in each time interval, thereby determining the star formation rate for each time bin \citep[e.g.][]{2005MNRAS.358..363C, 2006MNRAS.365...46O, 2014ApJ...783..110K, 2017ApJ...837..170L, 2018ApJ...861...13C, 2019ApJ...876....3L}.
We define logarithmically spaced time bins up to 0.63 Gyr, after which the bins are logarithmically spaced in redshift (excluding the first and last bins) to ensure a fair distribution of both young and old stellar population bins.
We adopt a total of 12 bins with the lookback time intervals
$t_{\mathrm{intervals}}=[0.0, 10.0, 22.9, 52.5, 120.2, 275.4, 631.0, 1.056\times10^3, 2.230\times10^3, 4.361\times10^3, 7.417\times10^3, 10.390\times10^3, 13.182\times10^3]$\,Myr.
\citet{2019ApJ...876....3L} investigate how varying the number of bins (up to 14) in a non-parametric fit affects the sensitivity of the results.
They find no clear trends, suggesting that their results are not significantly impacted by choice of bin number.

To constrain the SFR in each bin, we use the Student's-t distribution as priors as it has heavier tails than a normal distribution.
This permits sharp transitions in the SFR, such as those that might occur during quenching.
This approach follows the parameterisation adopted by \citet{2019ApJ...876....3L}.
This is normalized to fit for a total galaxy stellar mass formed between $10^5-10^{14}$\,M$_\odot$.
The metallicity for each spectral and IMF model variant is fixed at 70\% Solar ($Z=0.014$).
To investigate the fixed metallicity assumption, we conduct two additional tests using the default v2.2.1 BPASS framework when fitting.
The first test allows the metallicity to vary as a free parameter, fitting between $Z=0.0005-2\,Z_\odot$.
The other incorporates an evolving metallicity grid. This follows the evolution of the mass-metallicity relation presented by \citet{2015ApJ...800...20G}, assuming a typical galaxy mass of $\log(M/\mathrm{M}_\odot)=10$, interpolated onto the same time steps as the BPASS SPS models.

Dust attenuation is fitted for using the same \citet{2018ApJ...859...11S} laws for star-forming and quiescent galaxies that were used to generate the mock spectra.
Obtaining star formation rate and mass for a galaxy requires fitting and is thus not available to categorise a galaxy at this stage.
Therefore, we use colour information to determine which attenuation law to use when fitting.
We adopt the formalism of \citet{2001AJ....122.1861S} which separates SDSS galaxies using the \textit{u} and \textit{r} filters, where galaxies with $u-r>2.22$ are classified as quiescent and the rest are star-forming.
Both colour cut and sSFR agree on class categorisation for most simulated galaxies in EAGLE.
The amount of attenuation was allowed to vary between $A_V=0-5$\,mags.

Fitting spectra requires three additional variables to be constrained: instrument calibration issues, velocity dispersion, and noise.
Instrument calibration accounts for mismatches between spectroscopy, photometry and template models (i.e. inaccurate relative flux calibration as a function of wavelength).
When testing this parameter, we found it to mimic the behaviour of the attenuation law, resulting in inaccurate extinction values being derived.
We generated our mock spectra directly from templates, meaning there are no calibration issues and thus, we fixed this value to be flat.
The velocity dispersion was fitted for between 1 and 1000\,km/s.
Note that no intrinsic velocity dispersion was assumed and so this will be dominated by the SDSS spectral and BPASS template resolutions.
The stellar template libraries making up BPASS have different resolutions, meaning the apparent integrated light velocity dispersion can change with age.
This parameter must also be constrained during blind survey fitting, which this work is replicating.
Noise is accounted for by fitting for a multiplicative factor between 1 and 10 assuming white scaled noise.
Both the velocity dispersion and noise use logarithmic priors.

\section{Variations in Spectral Library} \label{sec:spec_lib_impact}

\subsection{Inferred Galaxy Properties} \label{sec:spec_lib_res}
\begin{figure}
    \centering
    \includegraphics[width=\columnwidth]{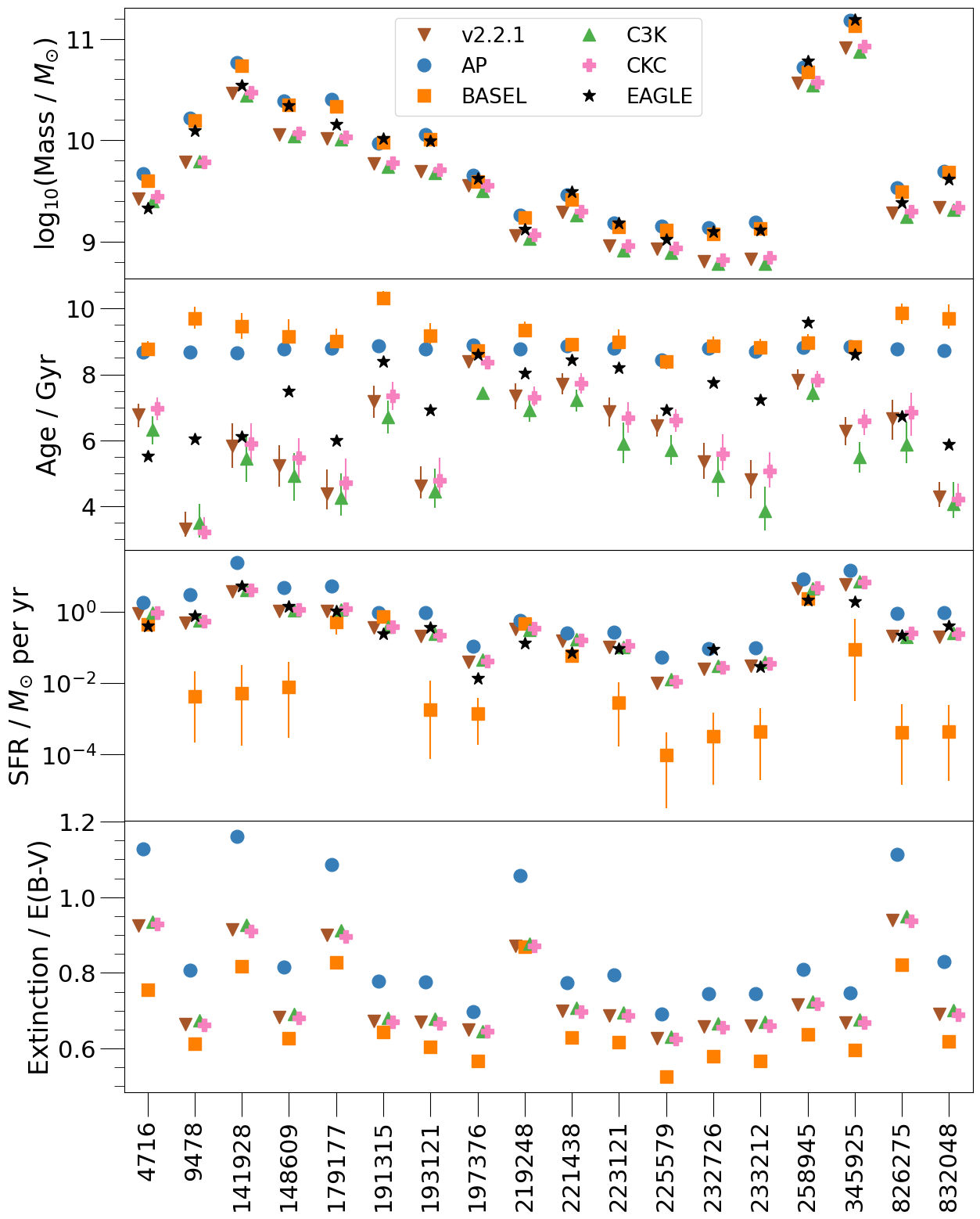} 
    \caption[]{Derived stellar mass, age, star formation rate, and extinction for each simulated galaxy, as inferred from spectral fits using SPS templates generated with different stellar spectral libraries.
    The brown inverted triangles represent the v2.2.1 template, blue circles for AP, orange squares for BaSeL, green upright triangles for C3K, and pink crosses for the CKC grid.
    EAGLE catalogue values are shown as black stars.}
    \label{fig:atmo_param_comp}
\end{figure}

The inferred parameters for each simulated galaxy when fitting using the AP, BaSel, C3K and CKC spectral libraries are shown in Figure~\ref{fig:atmo_param_comp}.
For comparison, the EAGLE catalogue values for mass, age and SFR are shown as black stars.
Extinction values from EAGLE are not included, as these were fixed during generation of mock observations and not taken from the catalogue.
The level of attenuation applied depends on whether the galaxy was classified as quiescent or star-forming.

Considering first the results from fitting using v2.2.1, we expect to recover the input values determined by EAGLE, with any remaining uncertainty due to the mock observation and fitting process rather than the stellar population properties.
The SFRs inferred from these default v2.2.1 fits are broadly consistent with those input from the EAGLE catalogue, while mass and age are generally underestimated during the fitting process.
Younger galaxies emit more flux per unit mass than older galaxies.
As a result, a lower inferred age requires a lower derived total stellar mass to generate the same observed luminosity.
This degeneracy between mass and age means that underestimating one often results in the underestimation of the other.

The underestimation in inferred galaxy properties may arise due to inherent differences between the EAGLE, BPASS and \bagpipes\ codes.
For example, the EAGLE simulation employs single-star evolution models to compute stellar feedback and chemical enrichment, whereas BPASS incorporates binary stellar evolution, introducing fundamental differences in the underlying stellar population assumptions.

Another contributing factor is the difference in how parameters are weighted.
The EAGLE catalogue and mock observations are constructed using mass-weighted quantities, while SED fitting infers luminosity-weighted properties.
Mass-weighting gives more constraining power to the  least massive stars in a stellar population, which are often older and dimmer, whereas luminosity-weighting emphasises the brightest, typically younger parts of a stellar population that dominate a galaxy's observed spectrum.
As a result, mass-weighted quantities tend to yield older ages and higher stellar masses, while luminosity-weighted estimates are biased towards younger ages and lower masses.
To isolate the effects of stellar population and spectral synthesis assumptions on the inferred parameters from those intrinsic to fitting, and reduce external sources of uncertainty, further comparisons will use the v2.2.1 default BPASS framework as a reference, rather than the EAGLE catalogue values.

\begin{figure}
    \centering
    \includegraphics[width=\columnwidth]{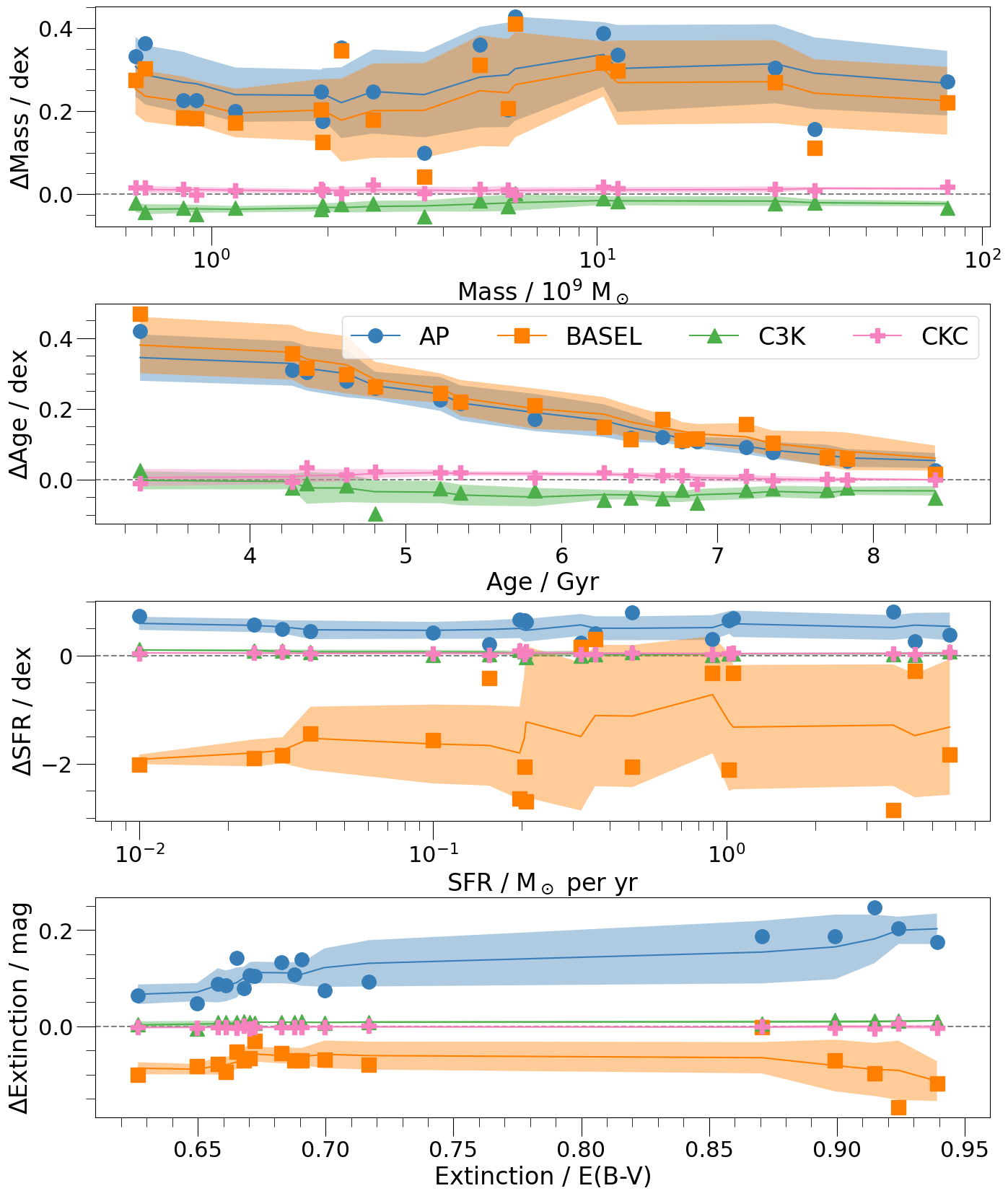} 
    \caption[]{Inferred galaxy property difference between the v2.2.1 BPASS framework and all spectral library variations, plotted as a function of the values obtained using the v2.2.1 framework.
    The AP, BaSeL, C3K and CKC spectral libraries are represented by blue circles, orange squares, green triangles, and pink crosses, respectively.
    Each data points corresponds to a single simulated galaxy fit, while solid lines show the running average over 6 data points with the shaded region indicating the one-sigma uncertainty.
    Error ranges on individual galaxies are not included for clarity.}
    \label{fig:uncert_atmo_comp}
\end{figure}

Fig.~\ref{fig:uncert_atmo_comp} shows the differences in derived stellar mass, age, SFR and extinction between reference values from the v2.2.1 BPASS framework and the other spectral libraries tested.
The mean offset in derived property values for all spectral libraries is summarised in Table~\ref{tab:atmo_offsets}.
The results inferred fitting using different spectral libraries fall into two distinct categories: those that yield results consistent with the v2.2.1 framework (C3K and CKC), and those that show significant disagreement (AP and BaSeL).
The discrepancies in the latter group often exceed $1 \sigma$, and in some cases even $3 \sigma$.
Fits obtained using both the AP and BaSeL templates systematically infer higher stellar masses and older ages across all galaxies.
The mass offset remains roughly constant across the full range of galaxy masses, while the age discrepancy decreases for older galaxies.
In terms of SFR and extinction, fits using the AP stellar library SPS model consistently predict higher values compared to v2.2.1, while those using BaSeL yield lower values for both parameters relative to the other libraries.

The v2.2.1 BPASS framework is built using the CKC library for the spectra of main sequence and giant stars, and C3K is an updated version of the CKC library.
As a result, they are expected to show good agreement with each other, but discrepancies could arise.
During the synthesis of an integrated light spectrum from a population, each stellar spectral template must be normalised such that the total bolometric luminosity for each spectrum is 1\,L$_\odot$.
However, the wavelength coverage of spectral templates in the C3K grid is narrower than that of CKC (only extends to 20\,000 instead of 100\,000\,\AA). As a result, the C3K spectra at fixed optical magnitude would have a higher implied bolometric luminosity without an additional normalisation step. This would result in lower inferred galaxy masses for a given observation.
To address this, \citet{2023MNRAS.521.4995B} implemented an additional normalisation step to match the C3K template to the CKC spectra.
If this step was imperfect, lower inferred galaxy masses could result, as suggested in this work (see Table~\ref{tab:atmo_offsets}). However, the corrections are typically very small, and should not dominate the overall uncertainty in this work. Hence the  discrepancies found are likely not due solely to our methodology.

\subsection{Model-Induced Variations in Galaxy Classification} \label{sec:diss_spec_lib}

\begin{table*}
    \begin{center}
    \caption[]{Mean differences in inferred galaxy properties for each stellar spectral library relative to those inferred by the default BPASS v2.2.1 fits.}
    \label{tab:atmo_offsets}
    \begin{tabular}{l | cccc}
        \hline
        Model & Stellar Mass / dex & Age / dex & SFR / dex & Extinction / mag \\
        \hline
        AP & $0.27 \pm 0.09$ & $0.17 \pm 0.11$ & $0.52 \pm 0.19$ & $0.13 \pm 0.05$ \\
        BaSeL & $0.23 \pm 0.09$ & $0.19 \pm 0.11$ & $-1.4 \pm 1.0$ & $-0.08 \pm 0.03$ \\
        C3K & $-0.03 \pm 0.01$ & $-0.04 \pm 0.03$ & $0.05 \pm 0.04$ & $0.008 \pm 0.004$ \\
        CKC & $0.011 \pm 0.007$ & $0.008 \pm 0.012$ & $0.04 \pm 0.02$ & $-0.001 \pm 0.002$ \\
        \hline
    \end{tabular}
    \end{center}
\end{table*}

Fitting using the AP and BaSeL models results in contrasting galaxy properties relative to the other spectral libraries.
BaSeL is the lowest resolution spectral library, which has a major influence on the results.
Low resolution spectra do not resolve line features, causing the wavelength-averaged continuum normalisation to decrease, especially at bluer wavelengths which reduces the relative amount of UV flux.
This results in BaSeL templates having a weaker 4000\,\AA\ Balmer break at late ages compared to other models \citep[as seen in Fig.~4 of][]{2023MNRAS.521.4995B}.
An older galaxy is thus required to generate the same relative Balmer break, which causes inferred ages to increase relative to the reference v2.2.1 BPASS framework.
This results in increased derived stellar masses to generate the same overall luminosity.

Since the BaSeL model cannot resolve UV line absorption features and an older population can mimic the spectral shape of an attenuated young population, the latter is not selected in the fits.
As a result, inferred SFRs are lower relative to other spectral libraries, and decreased extinction values are derived due to naturally redder spectra.
This variation in SFRs between inferred results obtained using different spectral libraries for fitting is significant, with BaSeL templates implying values which are up to three orders of magnitude lower.

Along with the increased stellar masses, the inferred sSFR also decreases significantly, leading to categorisation of galaxies as quiescent rather than star-forming (as would be the case when using the other spectral libraries).
As a result, the use of BaSeL templates for fitting can artificially inflate the quiescent galaxy population, distorting evolutionary pathways inferred from galaxy demographics across cosmic time.
We conclude that the low resolution of the BaSeL spectra renders them unsuitable for fitting to the high-quality simulated galaxies used in this study, as they fail to capture key spectral features necessary for robust parameter inference.

The AP stellar template library has a higher resolving power ($R\sim30\,000$), allowing for line features to be resolved and this enables SPS models to match the reference v2.2.1 BPASS framework in resolution (these are stepped down to 1\AA\ resolution in synthesis).
In this case, differences arise due to the spectral parameter grid used and the resulting stellar atmosphere templates derived.
For example, the temperature limit of this grid for cool stars is 3500\,K (compared to 2500\,K for CKC and C3K), meaning that all lower temperature stellar evolution models are assigned to the limit, making the population appear hotter and slightly bluer than it should be.
A similar issue occurs for the gravity grid, where the AP stellar template grid only extends down to $\log(g/\mathrm{cgs})=0$ rather than $\log(g/\mathrm{cgs})\sim-1$ (as in the other libraries).

Another difference can be found in line absorption. For the same age, AP templates have a stronger H$\beta$ absorption feature compared to CKC \citep[as seen in Fig.~6 of][]{2023MNRAS.521.4995B}.
This results in older inferred ages to match absorption features when fitting using models built with AP.
Old stellar populations (age above $\sim1$\,Gyr) inevitably contain a significant number of red stars which contribute to the overall spectrum.
\citet{2023MNRAS.521.4995B} show in the bottom panel of their Fig.~1 that cool star spectra differ dramatically between the CKC and AP models due to increased molecular absorption at bluer wavelengths in the AP library, yielding redder spectra at late ages.
The balance between these effects means increased SFRs are required to produce a young stellar population which matches the optical spectral line features, while increased extinction is necessary to correct the continuum slope and match the overall redder colour.
This naturally results in a lower overall flux normalisation, increasing derived stellar mass.

In summary, stellar spectral template libraries have fundamental differences between them which produce variation in the  templates assigned to stars with a given set of physical conditions, impacting the parameter space derived from the integrated light of a stellar population.
Thus, the choosing of a spectral library must be justified based upon the assumptions in the grid and the sample being modelled.

\section{Variations in Initial Mass Function}
\label{sec:imf_everything}
\begin{figure}
    \centering
    \includegraphics[width=\columnwidth]{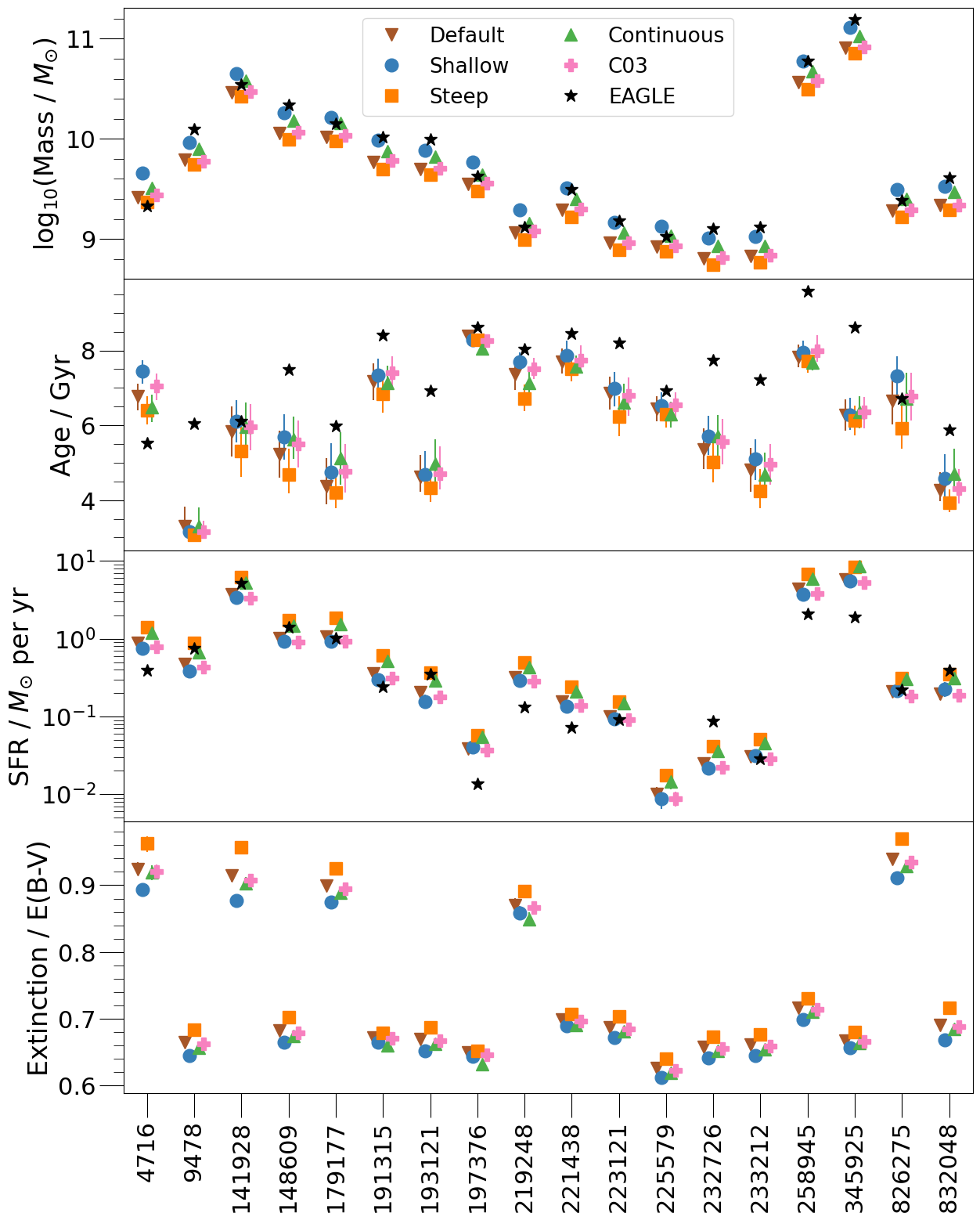} 
    \caption[]{Derived stellar mass, age, star formation rate, and extinction for each simulated galaxy, as inferred from spectral fits using SPS templates generated with different IMFs.
    The brown inverted triangles are for the default IMF prescription, blue circles for shallow, orange squares for steep, green upright triangles for continuous, and pink crosses for the Chabrier IMF prescription.
    EAGLE catalogue values are shown as black stars.}
    \label{fig:imf_param_comp}
\end{figure}

\begin{figure}
    \centering
    \includegraphics[width=\columnwidth]{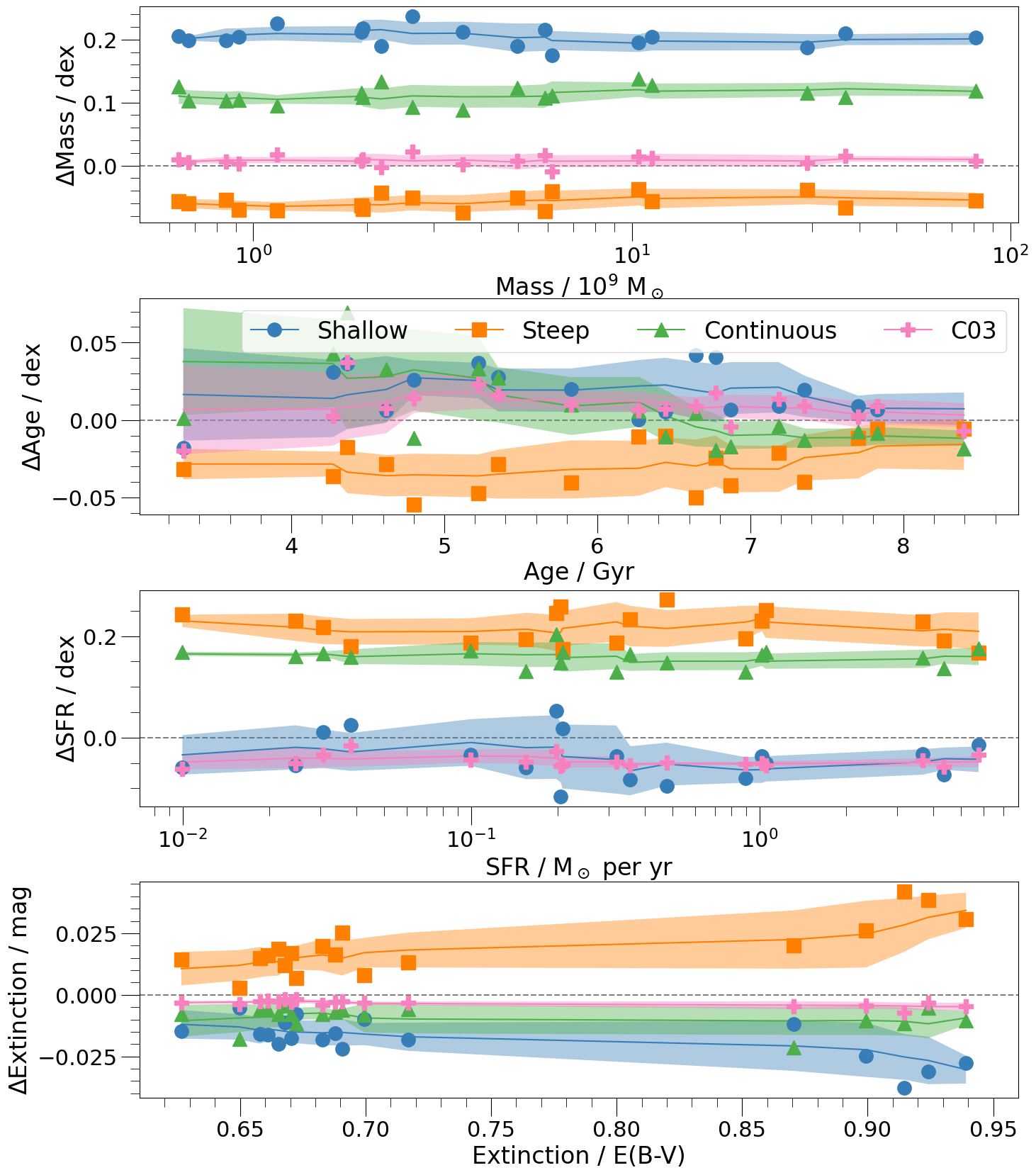} 
    \caption[]{Inferred galaxy property difference between the default IMF v2.2.1 BPASS framework and all other IMF prescriptions, plotted as a function of the values obtained using the default IMF.
    The shallow, steep, continuous and C03 IMF prescriptions are represented by blue circles, orange squares, green triangles, and pink crosses, respectively.
    Each data point corresponds to a single simulated galaxy fit, while solid lines show the running average over 6 data points with the shaded region showing the one-sigma uncertainty.
    Error ranges on individual galaxies are not included for clarity.}
    \label{fig:uncert_imf_comp}
\end{figure}

When modelling galaxies, the true IMF is an unknown.
Discrepancies will arise when fitting if conducted using an IMF that is not reflective of the true stellar population.
The properties inferred from fitting each mock galaxy using populations with different IMF assumptions is shown in Fig.~\ref{fig:imf_param_comp} and the difference to the reference BPASS v2.2.1 IMF is plotted in Fig.~\ref{fig:uncert_imf_comp}.
The mean offset in the derived parameter values for all IMF assumptions relative to the default IMF is summarised in Table~\ref{tab:imf_offsets}.
Notable differences in the derived parameters are observed, with the most significant variation occurring in the estimated stellar mass.

The C03 IMF prescription gives the best agreement with the default, only differing significantly in derived SFR.
The default IMF prescription is a broken-power law following the work of \citet{1993MNRAS.262..545K}, which utilised the stellar distribution found in the Milky Way disk to derive their IMF.
\citet{2003PASP..115..763C} define their IMF using stellar distributions in various parts of the Milky Way which includes the disk and globular clusters, finding IMF depends weakly on environment.
Thus, both the default and C03 prescriptions are based on the Milky Way IMF and their resulting forms are extremely similar.

The key differences are in the shape of the low-mass turnover (which is exponential in C03 rather than a power-law) and the slope of the high-mass power-law ($-2.35$ and $-2.30$ in the default and C03, respectively).
The latter is more influential, resulting in the C03 prescription having a slightly higher number of high-mass stars contributing to each spectrum.
The high-mass stars dominates the UV and blue-optical spectrum (see bottom panel of Fig.~\ref{fig:spec_comp_7}) and can dominate the overall fit.
However, the old stellar populations generally dominate the red-optical and near-infrared spectrum, and our EAGLE simulation objects are old galaxies, their red stars have greater influence on the spectral fit.
At late times, the C03 prescription generates slightly lower relative fluxes (see bottom panel of Fig.~\ref{fig:spec_comp_10}) for a given mass and star formation history.
This results in the C03 prescription deriving a lower SFR to match the blue part of the spectrum and a higher total stellar mass to produce enough low-mass stars to match the red emission.

The continuous \citet{1955ApJ...121..161S} IMF produces more low-mass stars relative to the default prescription as it is no longer a broken power-law and has a high-mass limit of 100\,M$_\odot$ instead of 300\,M$_\odot$.
Similar to the C03 IMF, it is the high-mass end which most affects the derived SFR.
Excluding stars in a population with stellar masses of $100-300$\,M$_\odot$ decreases the luminosity of the synthetic population for a fixed total mass, meaning a \citet{1955ApJ...121..161S} IMF leads to an inference of higher SFRs in order to generate the same overall galaxy luminosity.
The low-mass end of the IMF slope does not affect the luminosity in the optical, but contains most of the stellar mass.
Thus, a continuous IMF contains more mass for the same optical-NIR luminosity, increasing the galaxy stellar mass derived from fitting a set of mock observations, relative to the default IMF.

The difference between the shallow, default and steep IMFs is the upper-mass slope ($-2.00$, $-2.35$ and $-2.70$, respectively), altering the luminosity of model spectra. Fitting using these results in a variation in derived parameters, which arises from similar effects to those discussed in the context of the C03 IMF.
The shallow slope, having more massive-stars, generates brighter spectra at young stellar population ages for a given mass (see bottom panel of Fig.~\ref{fig:spec_comp_7}).
To generate the same UV luminosity as the simulated galaxies, fitting using the shallow IMF thus requires lower SFRs, while the steep IMF requires higher SFRs relative to the default prescription.
The steeper slopes have a higher fraction of low-mass stars per unit mass, creating a larger total flux normalisation at late ages.
This results in the steeper IMF slopes implying lower stellar masses.

The results presented above are strictly valid only for our selected combination of SDSS-like spectroscopy with VISTA-like NIR photometry. However, the trends and patterns seen here will systematically change the inferred galaxy parameters for any given set of observational data.

\begin{table*}
    \begin{center}
    \caption[]{Mean difference in inferred galaxy properties for each IMF prescription relative to those inferred by the default BPASS v2.2.1 fits.}
    \label{tab:imf_offsets}
    \begin{tabular}{l | cccc}
        \hline
        Model & Stellar Mass / dex & Age / dex & SFR / dex & Extinction / mag \\
        \hline
        Shallow & $0.21 \pm 0.01$ & $0.02 \pm 0.02$ & $-0.04 \pm 0.04$ & $-0.018 \pm 0.008$ \\
        Steep & $-0.06 \pm 0.01$ & $-0.03 \pm 0.02$ & $0.22 \pm 0.03$ & $0.02 \pm 0.01$ \\
        Continuous & $0.11 \pm 0.01$ & $0.01 \pm 0.02$ & $0.16 \pm 0.02$ & $-0.009 \pm 0.004$ \\
        C03 & $0.008 \pm 0.007$ & $0.01 \pm 0.01$ & $-0.05 \pm 0.01$ & $-0.003 \pm 0.001$ \\
        \hline
    \end{tabular}
    \end{center}
\end{table*}

Our analysis concludes IMF prescription generates an uncertainty in derived stellar mass, SFR and extinction of up to $\sim~0.2$, $0.2$ and $0.01$\,dex, respectively.
Similar results have been found in other studies \citep[e.g.][]{2023arXiv231018464W, 2024A&A...686A.138C, 2024ApJ...963...74W}.
Different analyses generally agree on derived SFR (sometimes with different levels of uncertainty) as this is always dominated by the young stellar populations, but have found mass offsets in either direction.
This is due to the choice of observational sample data fitted, choice of IMF prescriptions used when fitting and SED fitting methodology.

We focussed on relatively old, low redshift simulated galaxies. By contrast, the studies of \citet{2023arXiv231018464W}, \citet{2024A&A...686A.138C} and \citet{2024ApJ...963...74W} investigate samples of high-$z$ galaxies.
These are solely dominated by young stellar populations, where shallower IMFs have a lower mass-to-luminosity ratio in the rest-frame optical, causing lower inferred stellar masses.

In conclusion, it is clear that understanding the IMF (and its potential variations with cosmic time) is crucial, otherwise the results of SED fitting using an incorrect assumption can generate lasting systemic uncertainties in the results and affect any interpretation of galaxy evolution.
We present further validation of the IMF systematic uncertainties in Appendix~\ref{sec:tsc_imf_test_append}.

\section{Variation in Metallicity Prescription} \label{sec:met_impact}

\subsection{Investigating Metallicity Prior Prescriptions} \label{sec:met_results}
The analysis presented here has, up to this point, assumed a fixed metallicity  ($Z=0.014$).
This is common practice \citep[e.g.][]{2009ApJ...706.1364F, 2010A&A...523A..13P, 2011Ap&SS.331....1W, 2015MNRAS.447..786P, 2022ApJ...927..192Y, 2023A&A...669A..11P}, especially when using photometric data alone as metallicity is encoded primarily in narrow spectral line features and cannot be constrained in such cases \citep{2008A&A...491..713W}.
A fixed metallicity assumption generally holds true for local, massive galaxies which have undergone sufficient chemical enrichment to approach near-Solar values.
Metallicity is degenerate with other fitting parameters (e.g. stellar age, SFR), and so inaccurate metallicity assumptions can affect the fitting accuracy.

\begin{table*}
    \begin{center}
    \caption[]{Mean difference in inferred galaxy properties for the different metallicity assumption tests relative to the fixed-metallicity assumption.}
    \label{tab:met_offsets}
    \begin{tabular}{l | cccc}
        \hline
        Test & Stellar Mass / dex & Age / dex & SFR / dex & Extinction / mag \\
        \hline
        Variable & $0.01 \pm 0.04$ & $0.02 \pm 0.03$ & $0.03 \pm 0.20$ & $0.000 \pm 0.009$ \\
        Evolving & $0.02 \pm 0.04$ & $0.01 \pm 0.07$ & $-0.07 \pm 0.14$ & $-0.005 \pm 0.013$ \\
        \hline
    \end{tabular}
    \end{center}
\end{table*}

\begin{figure}
    \centering
    \includegraphics[width=\columnwidth]{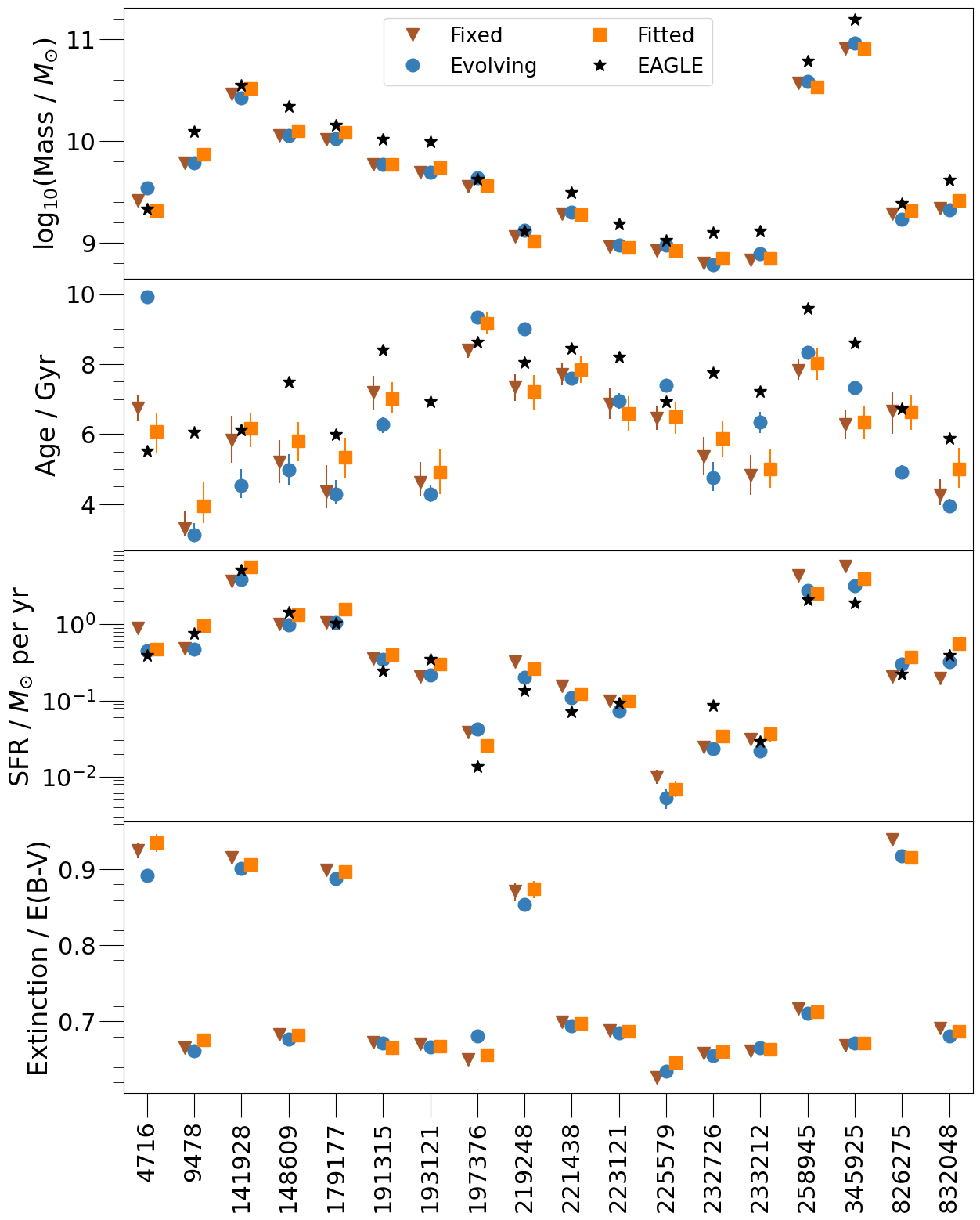} 
    \caption[]{Derived stellar mass, age, star formation rate and extinction from each metallicity assumption test.
    The brown inverted triangles represent the fixed-metallicity assumption ($Z=0.7$\,Z$_\odot$), blue circles for the evolving-metallicity grid, and orange squares for variable-metallicity.
    EAGLE catalogue values are included as black stars.}
    \label{fig:metal_param_comp}
\end{figure}

\begin{figure}
    \centering
    \includegraphics[width=\columnwidth]{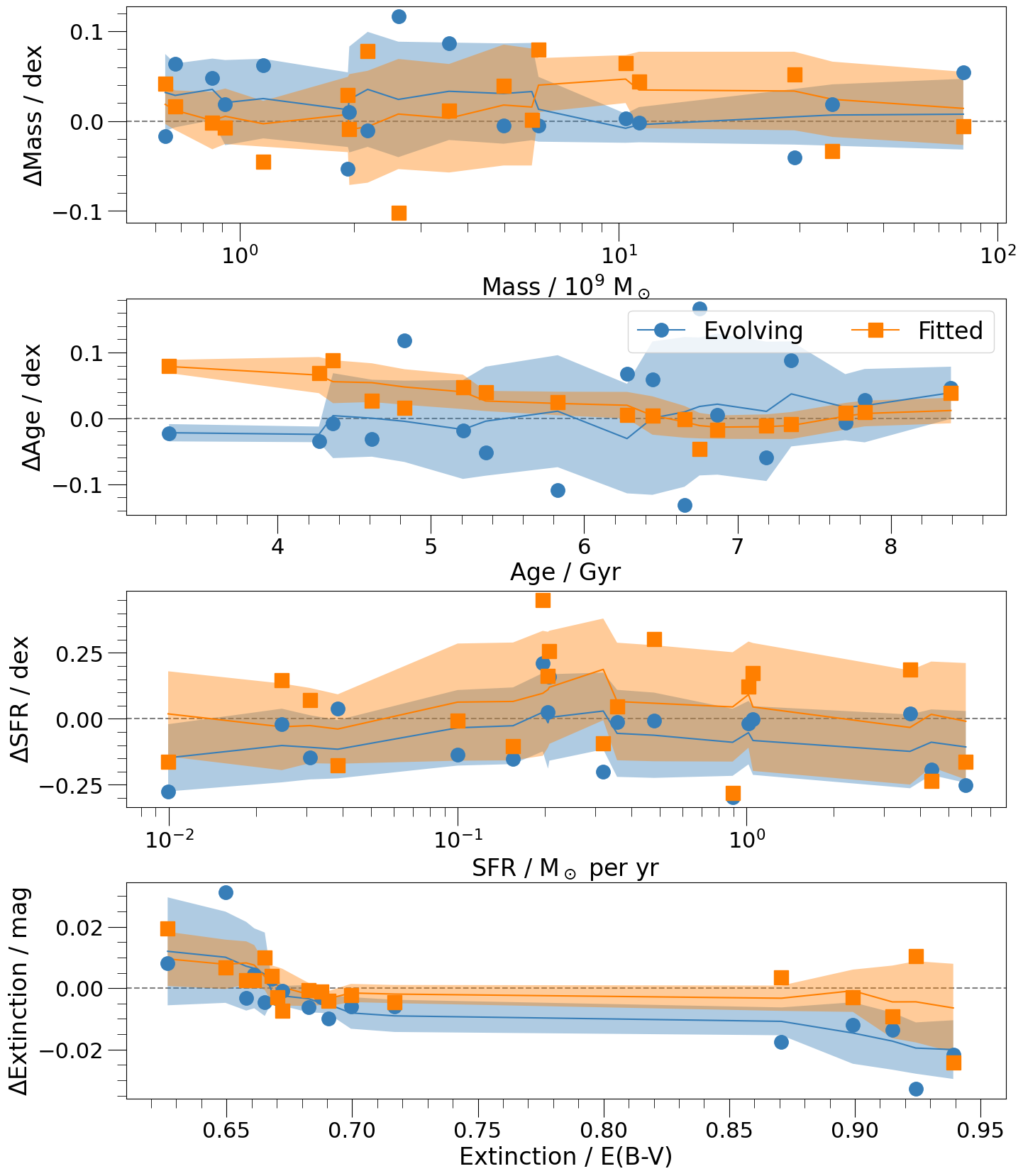} 
    \caption[]{Inferred galaxy property difference between the fixed-metallicity assumption run and the two other metallicity tests, plotted as a function of the fixed-metallicity run.
    The evolving-metallicity grid is displayed as blue circles and the variable-metallicity assumption as orange squares.
    Each data point corresponds to a single simulated galaxy fit, while solid lines show the running average over 6 data points with the shaded region indicating the one-sigma uncertainty.
    Error ranges on individual galaxies are not included for clarity.}
    \label{fig:uncert_met_comp}
\end{figure}

To explore the impact of this assumption, two additional tests were performed. We use the same mock observations as before (with EAGLE particle data defining their metallicity). In the first test, we fit these using synthetic stellar populations in which metallicity is treated as a free parameter. In the second test we use an SPS framework constructed such that metallicity varies as a function of star formation time following the mean cosmic chemical enrichment history as discussed in Section~\ref{sec:tsc_bagpipes}. 
We refer to these as the variable and evolving tests, respectively.
The former test is a standard approach for present-day SED fitting techniques \citep[e.g.][]{2017ApJ...837..170L, 2018MNRAS.480.4379C, 2019ApJ...879..116I, 2021ApJS..254...22J} while the latter is a recently adopted method. However by contrast with other work \citep[e.g.][]{2022MNRAS.511.5405B, 2024arXiv241017698B}, we do not permit a varying final metallicity. 
Fitting using an evolving metallicity should produce more realistic estimates of the true galaxy metallicity, but is rarely used in SED codes for reasons of computational resource and added degeneracy in the fit outcomes.
\bagpipes-derived parameters for each simulated galaxy as a result of these tests are shown in Fig.~\ref{fig:metal_param_comp}, including those from the fixed-metallicity run and the EAGLE catalogue values (a mass-weighted average of particle metallicities) for comparison.

Fig.~\ref{fig:uncert_met_comp} shows, for each galaxy, the difference in derived parameters between the two metallicity tests and the fixed-metallicity run, plotted as a function of the fixed-metallicity parameter values.
The mean difference across all galaxies in inferred  parameter values is presented in Table~\ref{tab:met_offsets}.
No strong trends are observed in any inferred property across the metallicity assumptions, with small scatter distributed randomly across the parameter space.

\begin{figure}
    \centering
    \includegraphics[width=\columnwidth]{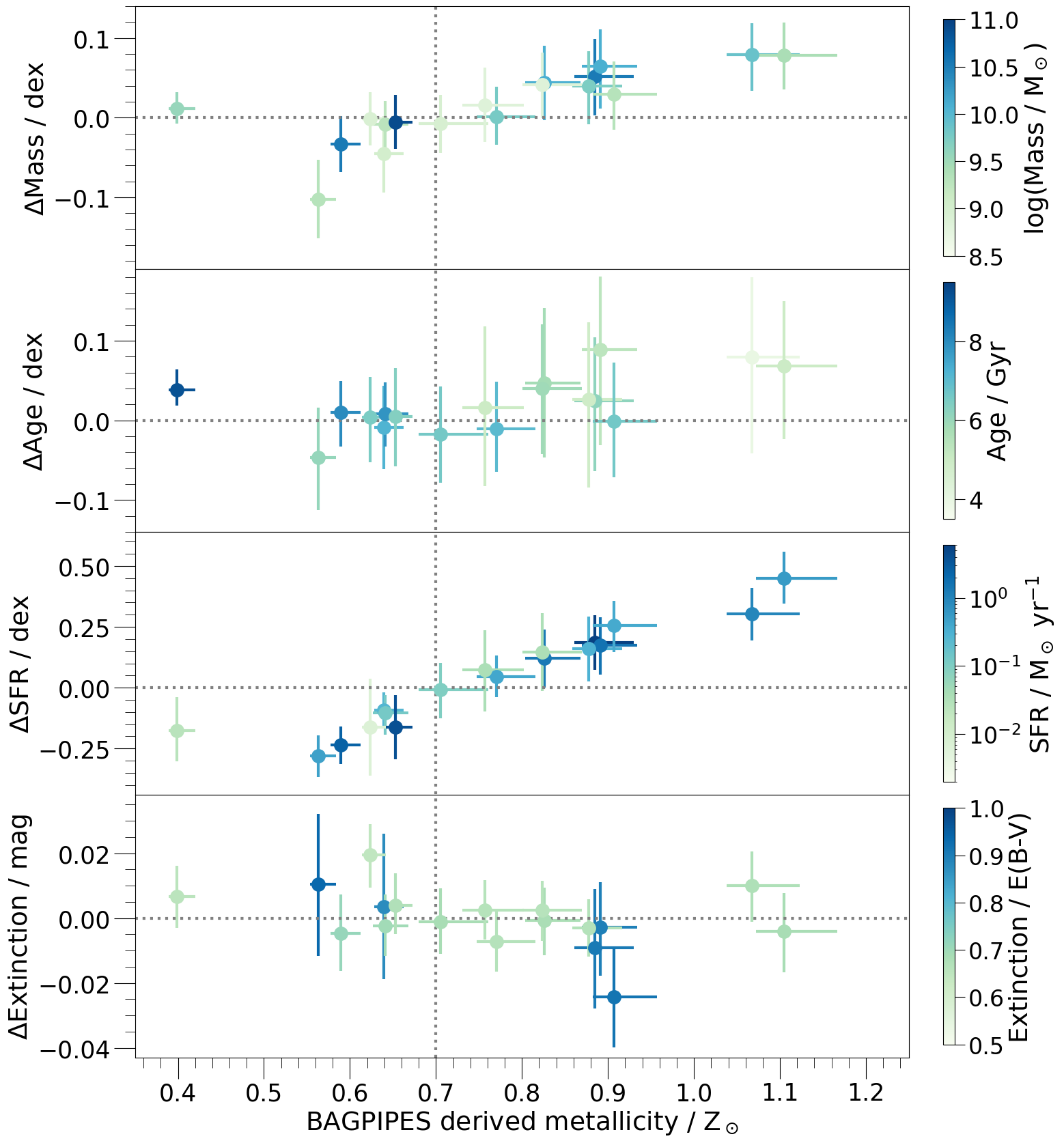} 
    \caption[]{Galaxy metallicity value derived from \bagpipes\ plotted against the inferred galaxy property difference between the variable- and fixed-metallicity assumption runs.
    Plotted, from top to bottom, are the parameters of stellar mass, mass-weighted age, star formation rate and extinction.
    Data points are coloured by the inferred parameter value for that subplot from the fixed-metallicity run, shown by the colour-bar at the right of each subplot.
    Inferred age and extinction show no trends with metallicity, while both mass and SFR have positive trends.
    The horizontal dotted line indicates no variation in derived parameters between the assumptions.
    The vertical dotted line is at $Z=0.014$, used in the fixed-metallicity run.}
    \label{fig:met_vs_params}
\end{figure}

To understand how inaccurate metallicity assumptions drive differences in the multidimensional output parameter space, the results from the variable- and fixed-metallicity runs are plotted as a function of the inferred metallicity from the variable run in Fig.~\ref{fig:met_vs_params}.
The colour of each data point represents the actual inferred value from the fixed run, as indicated by the colour bar beside each subplot.
The choice of metallicity prescription does not significantly affect the derived dust attenuation.
This implies that the estimated energy reprocessed into dust emission remains consistent regardless of the metallicity when at near-Solar values.
Similarly, metallicity prescriptions does not impact the derived stellar age.
However, we do find that older galaxies tend to have lower derived metallicities on average.
Older galaxies typically have stellar populations which formed in metal-poor environments, reducing the overall metallicity of the galaxy.
Both derived stellar mass and SFR exhibit a dependency on the derived metallicity value, with a higher inferred metallicity generally leading to systematically larger values for both parameters. 

\begin{figure}
    \centering
    \includegraphics[width=\columnwidth]{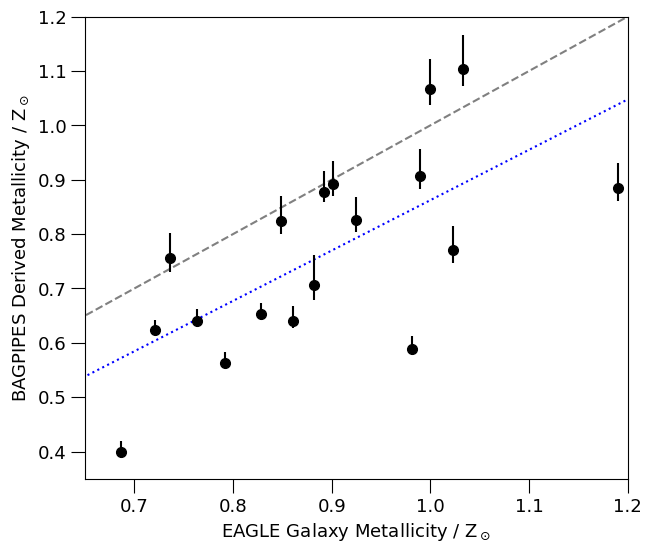} 
    \caption[]{Derived metallicity from \bagpipes\ against the EAGLE catalogue metallicity for each simulated galaxy.
    The grey-dashed line represents a one-to-one agreement between the two values while the blue-dotted line is the best-fit line to the data.
    \bagpipes\ systematically derives lower metallicities for each galaxy.}
    \label{fig:derived_fit_mets}
\end{figure}

No matter which metallicity prescription is adopted, derived stellar mass and age are consistently underestimated relative to the EAGLE catalogue values.
This indicates that the discrepancies are not driven by the choice of metallicity prescription.
We compare the metallicities derived from \bagpipes\ to those listed in the EAGLE catalogue in Fig.~\ref{fig:derived_fit_mets}.
\bagpipes\ fits typically underestimate metallicities by about 10\% of Solar, relative to the EAGLE values, while agreeing on the relative enrichment of different galaxies.
The posterior uncertainty quoted by the fitting algorithm for any individual galaxy is substantially smaller than the offset between its EAGLE and derived values, highlighting the challenge of constraining metallicity through SED fitting.
Since the EAGLE particle metallicity was used to generate the mock spectra, the expected recovered value should ideally match the catalogue value. However the latter is mass-weighted, rather than luminosity-weighted.
The observed discrepancy between input and derived value demonstrates the systematic differences that exist between the EAGLE, BPASS and \bagpipes\ codes.
In particular, the mass-weighted metallicities in the EAGLE catalogue reflect the total metal content and long-term chemical enrichment, while the particle-by-particle values implemented in construction of the mock observations yield luminosity-weighted totals, more similar to those derived by \bagpipes. These are more sensitive to the bright, short-lived massive stars.
As a result, a magnitude-limited, integrated light study will not probe the same aspects of a galaxy's chemical history as a resolved, mass-complete analysis such as that possible in simulation data.

\subsection{Exploring the Fundamental Plane} \label{sec:diss_met}
The galaxies studied in this sample are local systems that have undergone sufficient chemical evolution to become enriched with metals.
At near-Solar metallicities, differences in stellar emission spectra are relatively subtle.
Degeneracies between population level parameters can cause spectral variations that mimic the effects of metallicity.
This is particularly evident in the relation between derived metallicity and SFR shown in Fig.~\ref{fig:met_vs_params}.
Metallicity and SFR have opposing effects on a galaxy's SED.
Increasing the metallicity of a galaxy reddens a spectrum as metals contribute to line blanketing, which decreases UV flux relative to the IR.
In contrast, an increase in SFR means there are more young, UV luminous stars which make a spectrum bluer.
These competing effects balance against each other. If data cannot break the degeneracy between the effects, a false positive trend can be seen between metallicity and SFR. However real trends between these properties have been identified in previous studies \citep[e.g.][]{2012MNRAS.422..215Y, 2022A&A...666A.186D}. Indeed in very low mass galaxy samples negative metallicity-SFR relations \citep[e.g.][]{2012MNRAS.422..215Y} have been observed. The analysis here, suggests that such results must be interpreted cautiously.

A positive correlation between mass and metallicity is expected since it reflects the well-established, fundamental mass-metallicity relation in galaxies \citep[e.g.][]{2004ApJ...613..898T, 2006ApJ...647..970L, 2019ApJ...877....6B, 2024ApJ...966..228R, 2025arXiv250310800K}.
This relation is observed at all redshifts and reflects the fact that more massive galaxies have undergone more metal enrichment processes (i.e. supernova), lost less metals due to outflows, and are generally more efficient at accreting and converting gas into stars.
This genuine physical trend acts in the same way as the degeneracy in fitting that results from uncertain metallicities.
As a result, allowing metallicity to be fit as a free parameter may lead to an overestimate of the steepness of the mass-metallicity relation.

As Table \ref{tab:met_offsets} demonstrated, on average across galaxies spanning $\sim10^9-10^{11}$\,M$_\odot$ in stellar mass, the mean difference in derived stellar mass, age, SFR and extinction, between fitting for metallicity and fixing it (e.g. at $Z=0.014$) is less than the scatter in the fit values.
This suggests that for the widely-adopted practice of fitting stacked galaxy samples, rather than individual sources, assuming and fixing a representative metallicity is reasonable.
This representative metallicity must reflect the mass range and redshift of the sources since otherwise biases are introduced.
For example, adopting a derived metallicity of Z=0.012 implies a SFR and stellar mass that are $\sim0.2$ and $\sim0.05$\,dex lower, respectively, compared to fitting at $0.7$\,Z$_\odot$.
Conversely, selecting Z=0.022 yields values that are $\sim0.4$ and $\sim0.08$\,dex higher.
In cases where metallicity can be reliably constrained (i.e. from nebular emission or high-quality absorption line fitting), empirical values should be used rather than assuming a metallicity.

\citet{2024arXiv241017698B} studied a volume-complete sample of approximately 8,000 low redshift galaxies from the GAMA survey to investigate the impact of implementing a metallicity evolution prescription during fitting, compared to a fixed or constant metallicity.
They found that fixing metallicity introduces the largest bias in inferred galaxy properties, exceeding the effects of varying SFHs and SPS components such as isochrones, stellar spectral grids, and IMF.
In particular, they found that assuming a fixed (Solar) metallicity can lead to an underestimation of stellar mass by up to 0.27\,dex and an overestimation of SFRs by around 0.2\,dex, relative to a reference value defined by their preferred SED fitting technique.
These offsets are comparable in magnitude to those found in our work, compared to the known input values for individual galaxies. However, the results of our analyses differ in that  \citet{2024arXiv241017698B} observed an offset in the average values for their sample, whereas we find the average offsets to be very small. 
This difference likely arises because their sample includes galaxies with stellar masses extending down to $10^7$\,M$\odot$, whereas our sample is limited to galaxies with masses above $10^9$\,M$\odot$. The lowest mass galaxies would be expected to have the lowest metallicities and thus show still larger offsets than those determined in our fitting tests.
As shown in Fig.~6 of \citet{2024arXiv241017698B}, the effects of metallicity assumptions are most pronounced at these lowest stellar masses, where a fixed Solar metallicity assumption is no longer appropriate.

\section{Impact on Cosmic Mass Assembly History} \label{sec:mah_impact}
One key outcome of SED fitting in galaxy surveys is the cosmic mass assembly history, tracking the fraction of present-day ($z=0$) stellar mass formed prior to a given lookback time.
Analysis in previous sections have examined how SPS model assumptions influence the derived properties of individual galaxies; however, these assumptions also affect large-scale, global quantities that emerge from the combination of individual galaxy properties.

\begin{figure*}
    \centering
    \includegraphics[width=\textwidth]{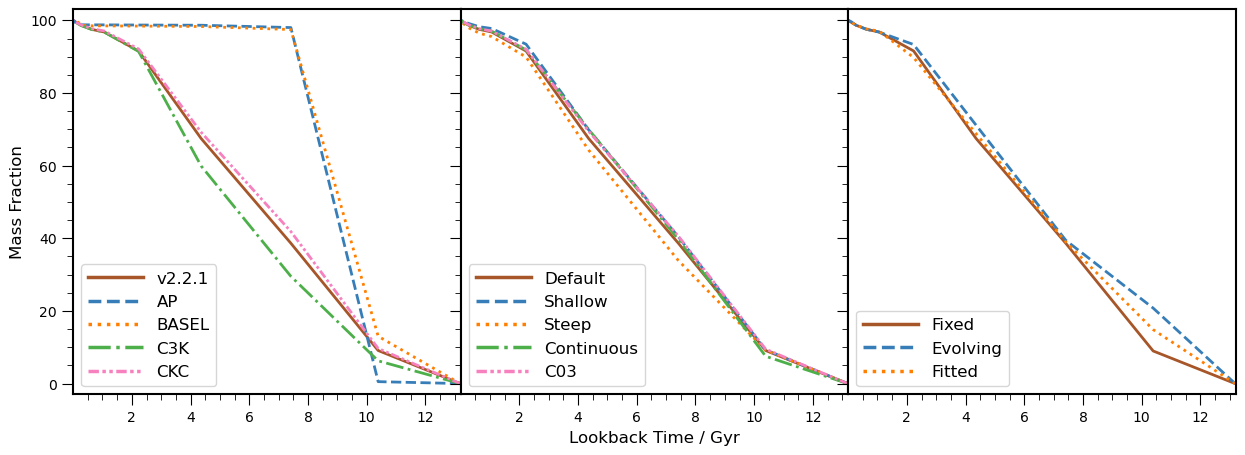} 
    \caption[]{Cumulative stellar mass fraction formed in all simulated galaxies fitted in this work, shown as a function of lookback time.
    The v2.2.1 BPASS framework with default IMF and fixed metallicity prescriptions highlight the comparison results plotted as the solid brown line in all panels.
    Left panel shows the results from fitting using the different spectral libraries of AP (dashed blue), BaSeL (dotted orange), C3K (dot-dashed green), and CKC (double dotted-dashed pink).
    Middle panel plots the results from fitting using the different IMF prescriptions of shallow (dashed blue), steep (dotted orange), continuous (dot-dashed green), and C03 (double dotted-dashed pink).
    Right panel shows the fitting results from the different metallicity tests of the evolving (dashed blue) and variable (dotted orange line) metallicity prescription.
    }
    \label{fig:mah_panel}
\end{figure*}

Fig.~\ref{fig:mah_panel} shows the cumulative stellar mass build-up over cosmic time for the full simulated galaxy sample. We show how this is influenced by presumed spectral library, IMF, and metallicity prescriptions used in the SED fitting algorithm in the left, middle, and right panels, respectively.
Among these, the IMF prescription has the smallest effect on the recovered mass assembly history.
The steep IMF assumption would result in a lower estimate of cumulative mass fraction formed (by approximately 4 percent) between lookback times of $\sim4$ and $8$\,Gyr relative to the default IMF.
In contrast, the shallow and C03 IMFs lead to a modest increase in estimated mass fraction formed in this epoch of $\sim2$ percent.
Even though 
the assembly histories reconstructed from the SED fits follow similar trends,
the final stellar mass formed by a lookback time of zero varies hugely between model results because each prescription alters the ratio in number between low-mass and high-mass stars.

The impact of varying metallicity prescriptions in the fitting is more significant, especially at early times.
The evolving metallicity grid leads to a faster initial build up being reconstructed, with up to a 12 percent increase in initial mass formed between the onset of star formation and a lookback time of 9\,Gyr.
After this point, all metallicity models predict the same assembly history to within $\sim3$ percent.
The largest discrepancies arise from the choice of stellar spectral library.
While the mass assembly history derived from the CKC model closely follows the v2.2.1 BPASS framework within 3 percent at all lookback times, the more recent C3K templates lead to a slower build-up of stellar mass being reconstructed, aligning with the v2.2.1 framework only within the last 2\,Gyr.
The C3K library gives a maximum deviation of 9 percent at a lookback time of $\sim7$\,Gyr relative to the v2.2.1 framework.

More extreme trends are seen with the AP and BaSeL libraries, both of which suggest that over 97 percent of total stellar mass in the galaxies being fitted was formed by a lookback time of $\sim7$\,Gyr.
In the AP model fits, this occurs almost entirely within a single time bin ($7.417<t<10.390$\,Gyr), while BaSeL model derived masses have a small fraction of star formation extending into the oldest bin.
Fitting using these spectral libraries would suggest that galaxies in the sample experience a rapid early star formation phase, followed by extended quiescence. This is known to be untrue given the EAGLE simulation star formation histories. 
When cross-referenced with the derived SFRs in Fig.~\ref{fig:atmo_param_comp}), fitting with BaSeL templates would indicate that most galaxies remain quiescent at the present day.
In contrast, AP fits show renewed star formation in some galaxies, yielding the highest present-day SFRs of any model considered. Again, the known properties of the EAGLE galaxies rule out this scenario.

As previously discussed in Section~\ref{sec:diss_spec_lib}, variations in the parameter grid coverage and spectral resolution of stellar spectra strongly affect derived galaxy parameters.
These same factors influence the inferred mass assembly history.
On average, each  time bin in the reconstructed mass assembly history, contains only $2-4$ BPASS SSP age bins.
In bins dominated by older stellar populations, where red supergiants, whose spectral energy distributions differ substantially between libraries, contribute significantly to the emission, the impact of uncertainties will be disproportionately large.
These spectral discrepancies are discussed in more detail by
\citet[][]{2023MNRAS.521.4995B}.

\citet{2024arXiv241017698B} also investigate how some variations in assumed SPS model components impact the derived cosmic star formation history (the integral of which traces the stellar mass assembly history).
Consistent with our findings, they show that the choice of IMF has the smallest effect on the recovered SFHs relative to other modelling assumptions, due to offsets in inferred stellar mass.
However, they do not find a strong sensitivity to the choice of spectral library, reporting minimal differences between fits using the C3K and BaSeL spectral libraries.
This contrasts with our findings, where we observe substantial differences in the resulting SFHs depending on the spectral library used, as illustrated in the left panel of Fig.~\ref{fig:mah_panel}.
One possible reason for this discrepancy lies in methodological differences between the studies, including assumptions regarding other input priors.
\citet{2024arXiv241017698B} fit only to photometric data, whereas our analysis also incorporates spectroscopic information, making our results more sensitive to detailed spectral features and to systematics introduced by the choice of spectral template library.
Consequently, the impact of incorrect SPS assumptions is likely to grow as future surveys provide deeper and higher-resolution spectroscopic datasets.

This analysis has demonstrated the relative uncertainty introduced by various SED fitting assumptions, both on small-scale and large-scale galaxy quantities.
We find a typical scatter of up to $\sim10$ percent in the cumulative stellar mass formed at early cosmic times (lookback time $>2$\,Gyr) reconstructed from SED-fit star formation histories.
These findings also underscore the importance of consistently modelling stellar populations and their prior assumptions when interpreting galaxy evolution from SED fitting.

It is  important to note that several input priors tested in our analysis have been found to correlate.
For example, metallicity uncertainties in fitting introduces biases in the derived IMF, and may also alter both the intrinsic IMF, with metal-poor systems favouring  top-heavy IMFs \citep[e.g.][]{2021MNRAS.508.4175C, 2024MNRAS.527.7306T}.
Consequently, varying multiple priors simultaneously would provide a more robust test of the assumptions involved in SED fitting and the resulting uncertainty in derived galaxy properties on both an individual and sample basis.

\section{Consequences for Surveys} \label{sec:tsc_consequences}
Astronomical surveys, such as SDSS \citep{2000AJ....120.1579Y, 2003MNRAS.341...33K}, Cosmic Evolution Survey \citep[COSMOS,][]{2016ApJS..224...24L, 2022ApJS..258...11W}, Galaxy and Mass Assembly, \citep[GAMA,][]{2011MNRAS.413..971D, 2015MNRAS.452.2087L, 2020MNRAS.498.5581B}, and Cosmic Assembly Near-infrared Deep Extragalactic Legacy Survey \citep[CANDELS,][]{2011ApJS..197...35G, 2011ApJS..197...36K, 2016MNRAS.462.4495H}, catalogue thousands of objects simultaneously and have been instrumental in reconstructing the stellar mass build-up and star formation histories of galaxies.
However, each such study typically adopts a single SED fitting methodology, with limited variation in the underlying SPS models or prior assumptions.
As a result, the reported uncertainties on derived galaxy properties often reflect only observational and instrumental errors, overlooking the substantial contribution from modelling choices.

\citet{2003MNRAS.341...33K}, who derived stellar masses and SFHs for SDSS galaxies, demonstrated that switching from a \citet{2001MNRAS.322..231K} IMF to a \citet{1955ApJ...121..161S} prescription effectively doubles the inferred stellar mass due to the latter's heavier weighting towards low-mass stars.
Importantly, the integrated light from galaxy stellar populations is insufficient to distinguish between the two IMF cases, as it is also unable to clearly distinguish between other model assumptions.
When fitting SEDs using a single prescription, one implicitly assumes that the physics encoded in that model sufficiently captures the real diversity of galaxies: an assumption often invalidated by observed variations in properties such as the IMF \citep[e.g.][]{2013ApJ...771...29G, 2020ApJ...899...87M, 2020ARA&A..58..577S}.

Our analysis has highlighted the extent to which stellar population model uncertainties propagate into errors on physical galaxy properties, even when using both photometric and spectroscopic observations to constrain the spectral energy distribution.
The implications are particularly critical for current and upcoming large-scale surveys, such as \textit{Roman} \citep{2015arXiv150303757S, 2019arXiv190205569A} and \textit{Euclid} \citep{2024arXiv240513491E}, which will primarily rely on photometric SED fitting for their largest samples.
The \textit{Euclid} science team have demonstrated that they will be able to study the galaxy main sequence at $0.9<z<1.82$ and SFRs as low as 0.1\,M$_\odot$\,yr$^{-1}$ in the Euclid Wide Survey and $\sim0.01$\,M$_\odot$\,yr$^{-1}$ in the Euclid Deep Survey \citep{2024arXiv240513491E}.
However, while these precision levels are within reach in terms of direct observational errors, the accuracy of inferred measurements remains limited by the underlying modelling assumptions: uncertainties that currently exceed the quoted measurement thresholds.

The \textit{Roman} science team has outlined synergies to complement missions such as Rubin, \textit{Euclid}, \textit{JWST} and \textit{Hubble}, leveraging broad wavelength coverage which should enable better SED modelling and more accurate determinations of SFR and stellar mass \citep{2015arXiv150303757S, 2019arXiv190205569A}.
The goal of such meta-analyses is precise and authoritative measurements of galaxy physical properties. The precision of these is often assumed to scale with the degree of photometric precision obtainable (i.e. to be observationally limited).
Our results demonstrate that this is unlikely to be true. Even spectroscopic and photometric data spanning the optical to near-infrared regime at high precision does not eliminate SED modelling uncertainties. Extending to the mid-infrared does not provide significant advantages in this respect since it would introduce uncertainties in the warm and hot gas emission spectrum \citep{2023MNRAS.525.5720J} and in AGN components. As a result we may well now be in a model-limited rather than observation-limited regime.

Despite the transformative datasets from future wide-field surveys uncertainties arising from assumptions baked into SPS models and their modelling uncertainties will become the dominant limiting factor in interpreting galaxy population properties such as stellar mass, age, SFH and thus mass assembly history.

Refining stellar population models depends on high-quality empirical data that remains sparse for key evolutionary stages and parameter regimes.
For instance, current spectral libraries diverge significantly in their treatment of red supergiants and massive stars at low metallicity \citep[e.g.][which will require better ultraviolet observations]{2023MNRAS.521.4995B}, and our knowledge of the initial mass function's potential variation across time and environment is still limited \citep[e.g.][]{2010ARA&A..48..339B}.
Similarly, binary stellar evolution, crucial for interpreting UV fluxes and high-mass stellar populations, requires better constraints on parameters like the binary fraction as a function of stellar mass and its metallicity dependence \citep[e.g.][]{2020MNRAS.495.4605S, 2020MNRAS.497.2201S}.
Efforts like BLOeM \citep{2025arXiv250212239B, 2025arXiv250202644P, 2025arXiv250321936V}, \textit{Gaia} \citep{2023A&A...675A..89D}, ULYSSES \citep[][]{2025ApJ...985..109R}, and ULYSSES X-shooter \citep{2023A&A...675A.154V} are helping to address this, with future breakthroughs likely to come from studying spatially resolved stellar populations in the Local Group using instruments such as \textit{JWST} or Extremely Large Telescope \citep[ELT, e.g. HARMONI,][]{2021Msngr.182....7T}.

Until these improvements are implemented \citep[e.g.][]{2025MNRAS.537.2433B, 2025MNRAS.537.2782Z}, uncertainties in stellar population modelling must be explicitly folded into the error budgets of inferred galaxy property estimates that result from using them.
Doing so is essential not only for accurately interpreting galaxy evolution, but also for ensuring the scientific return of the next generation of surveys.
The analysis in this work provides a critical step in that direction, quantifying the systematic uncertainties tied to SPS model choices in a scenario where the true values are known, and enabling a more realistic understanding of the limits of our inferences.

\section{Conclusions} \label{sec:tsc_concs}
This work has explored the impact of key assumptions in the construction of stellar population synthesis models on the derivation of galaxy properties and the reconstructed cosmic mass assembly history.
Our analysis utilised a sample of 18 simulated galaxies randomly selected from the EAGLE simulation project, for which mock SDSS spectra and VISTA photometry were generated.
Galaxy parameters were inferred using the SED fitting tool \bagpipes\ \citep{2018MNRAS.480.4379C} using a suite of BPASS synthetic stellar populations \citep{2017PASA...34...58E, 2018MNRAS.479...75S} devised with varying stellar spectral library and IMF prescriptions.
Each fit involves identical stellar astrophysics, population synthesis code and fitting algorithm priors while only the key assumptions of spectral library, IMF, or metallicity prior was altered.
Our main findings are summarised as follows:

\begin{enumerate}
    \item The variation between stellar spectral libraries introduces significant uncertainties in derived galaxy parameters due to inherent differences in their resolution and parameter coverage.
    Across the spectral templates tested, absolute uncertainties in stellar mass, age, SFR and extinction are $0.27\pm0.09$, $0.19\pm0.11$, $1.4\pm1.0$\,dex and $0.13\pm0.05$\,mag, respectively.
    These uncertainties can be minimised by carefully selecting a spectral template that is appropriate for the sample investigated.

    \item Varying the IMF prescription can introduce systematic biases. An IMF with a steeper upper-mass slope ($\alpha=-2.70$ versus $-2.35$) results in systematic offsets of $-0.06\pm0.01$ and $0.22\pm0.03$\,dex in stellar mass and SFR, respectively, while a shallower upper-mass slope ($\alpha=-2.00$) yields $0.21\pm0.01$ and $-0.04\pm0.04$\,dex.
    The derived age remains largely unaffected.
    Although the magnitude of these shifts is consistent with previous studies, the direction of the systematic offsets can differ.
    This is caused by the galaxy observations being fitted, where fitting a young stellar population with a shallow IMF results in lower inferred masses, while older populations yield higher masses.

    \item Fixing the metallicity when fitting individual galaxies introduces systematic offsets in mass and SFR.
    For example, fixing metallicity at $Z=0.014$ leads to offsets from the true values of $\sim 0.4$ and $\sim 0.08$\,dex in stellar mass and SFR, respectively.
    However, for stacked local galaxy samples with individual galaxy masses $>10^9$\,M$_\odot$, the metallicity prescription has negligible impact on derived properties.
    It is therefore reasonable to fix the metallicity in such cases, aligning with the expected cosmic chemical enrichment history.

    \item The \bagpipes-derived parameters of stellar mass, age and metallicity are systematically lower than those provided for the same galaxy in the EAGLE catalogue.
    This discrepancy arises from the difference in weighting, where EAGLE values are mass-weighted, while SED fitting is inherently luminosity-weighted.

    \item The fraction of the total stellar mass  formed during early galaxy formation (first $\sim10$\,Gyr), can differ by up to 12 percent depending on the SPS assumptions used.
    The choice of stellar spectral library has the most significant impact, in some cases shifting galaxies from being interpreted as experiencing an extended star-forming history to predominantly quiescent. These interpretations are at odds with the true galaxy properties.

    \item We conclude that the inference of galaxy parameters is strongly influenced by model assumptions which are not widely included in the error budget of the estimated properties. Given the advent of large, wide-field and sensitive photometric and spectroscopic surveys, we may now be in a regime dominated by modelling uncertainties rather than observational limitations.
\end{enumerate}

This study does not seek to determine the correct set of modelling assumptions to apply, but rather to quantify the uncertainties that arise by not correctly accounting for them.
While formal observational uncertainties in stellar mass and SFR are generally quoted as $<0.2$\,dex \citep[e.g.][]{2009ApJS..185....1T, 2010ApJ...722....1T, 2014RvMP...86...47C}, our findings suggest that uncertainties introduced by modelling assumptions can be equal to or exceed these values.
For example, differences between population synthesis models have been shown to introduce $0.2-0.3$\, dex of uncertainty \citep[e.g.][]{2006ApJ...652...85M, 2010ApJ...709..644I, 2024arXiv241017698B} and in some cases up to $\sim0.6$\,dex \citep{2009ApJ...699..486C}, while the choice of SED fitting code (when using an identical SPS framework) contributes 0.14 and 0.28\,dex of uncertainty for mass and SFR, respectively \citep{2022arXiv221201915P}.
These are in addition to the effects due to stellar spectral library and IMF prescription variations highlighted in this study and others \citep[][]{2023arXiv231018464W, 2024A&A...686A.138C, 2024ApJ...963...74W, 2024arXiv241017698B}.
Similar discrepancies have been identified when fitting individual stellar clusters \citep[e.g.][]{2025arXiv250602123M}, highlighting that these uncertainties are not limited to global galaxy properties but arise from model-dependent choices at the stellar population level.
Thus, despite improvements in data quality, uncertainties in derived galaxy properties remain large.

Constraining these modelling assumptions requires high-quality data, observing and modelling individual stars.
This has been limited to our Galaxy and a handful of nearby systems.
These observations are deficient in high-mass or low-metallicity stars which are either too faint to observe or only found in the distant Universe where they cannot be individually resolved.
Quantifying modelling uncertainties is especially critical as  observatories such as \textit{JWST} continue to explore previously uncharted and largely uncalibrated regimes.
This work has underscored the significant role of stellar spectral libraries and IMF prescriptions in shaping inferred galaxy properties, contributing to efforts to establish a robust error budget in SED-derived measurements.
Such efforts are essential to accurately probe the physical processes that govern galaxy formation and evolution.

\section*{Acknowledgements}
ERS, CMB and GTJ acknowledge support from the UK Science and Technology Facilities Council (STFC) through consolidated grants ST/T000406/1, ST/X001121/1, and a doctoral studentship.
This work made use of the University of Warwick Scientific Computing Research Technology Platform (SCRTP) and Astropy\footnote{\url{https://www.astropy.org/}}, a community-developed core Python package for Astronomy \citep{astropy:2013,astropy:2018}. 
We acknowledge the Virgo Consortium for making their simulation data available.
The EAGLE simulations were performed using the DiRAC-2 facility at Durham, managed by the ICC, and the PRACE facility Curie based in France at TGCC, CEA, Bruyèresle-Châtel.

\section*{Data Availability}
All simulation data products underlying these results are publicly available.
Specific results of fits will be made available via the BPASS website at \url{www.warwick.ac.uk/bpass} or \url{bpass.auckland.ac.nz}, or by reasonable request to the first author.



\bibliographystyle{mnras}
\bibliography{references}


\appendix

\section{Investigating uncertainty arising from evolutionary tracks} \label{sec:evo_tracks}
The common parameterisation of stellar evolutionary tracks are isochrones: a line depicting a stars position on the Hertzsprung-Russell diagram at each stage of its lifetime, determining stellar properties such as radius, luminosity, and surface gravity.
As with the other SPS framework components, various isochrone tables exist in literature \citep[e.g.][]{1992A&AS...96..269S, 1994A&AS..106..275B, 2000A&AS..141..371G, 2000A&A...361..101M, 2004ApJ...612..168P, 2007AJ....133..468C, 2008A&A...482..883M, 2012MNRAS.427..127B, 2018MNRAS.476..496F, 2022MNRAS.509.5197S, 2024MNRAS.527.2065P}.
Recent work has highlighted that isochrone selection introduces similar uncertainties in derived galaxy parameters relative to the SPS model stellar library and IMF parameterisation components \citep[e.g.][]{2024arXiv241017698B}.

The use of isochrones are invalid when binary evolutionary pathways are incorporated.
Binary interactions smear a single star's isochrone track into an isocontour: a broad region in the Hertzsprung-Russell diagram representing the varied outcomes of mass transfer, mergers, and other binary processes.
The range of possible evolutionary tracks depends on the binary interaction parameters assigned, such as stellar mass ratio and binary period.
The relative weighting of these binary parameters when generating SPS models remains a major uncertainty, which propagates through the choice of evolutionary tracks and perturbs the inferred galaxy parameters.
The treatment of binary parameters in BPASS is detailed at length in \citet{2017PASA...34...58E}.

The BPASS team has explored the impact SPS models using single-only evolutionary tracks have relative to binary-incorporated models \citep[e.g.][]{2022MNRAS.514.5706J}.
The work by \citet{2018MNRAS.477..904X} investigates the nebular emission from H{\sc ii} regions around young stellar populations.
They highlight that while emission lines disappear from single-only track spectra by $\sim10$\,Myr, binary-incorporated spectra have increased line emission which persists to $>100$\,Myr.
The BPASS team also looks into variations in their evolutionary tracks.
\citet{2025MNRAS.537.2433B} investigates how alpha-enhancement affects the evolutionary properties of stars in which they find strong dependences, such as in the main sequence lifetime which varies by up to $0.3$\,dex.
There have also been studies into the binary parameter distributions.
\citet{2023MNRAS.522.4430S} highlights that stochastic sampling of both IMF and the binary parameters has a significant impact on inferred properties, especially in lower mass stellar populations (e.g. $\mathrm{M}<10^6\,\mathrm{M_\odot}$).

Current BPASS releases have only one underlying evolutionary track variant, meaning a full investigation into the impact of evolutionary tracks is not possible.
However, BPASS does release both single track only SPS models and ones which incorporate binary evolutionary tracks.
Thus, we present an analysis into the uncertainties resulting from incorporating binary evolution into SPS models as a basis for understanding the uncertainties that can arise when fitting using different evolutionary tracks.
BPASS recognises that the Universe is full of binaries \citep[e.g.][]{2012Sci...337..444S, 2017ApJS..230...15M} and thus populations containing only single-stars (rather than a combination of binaries with singles) are unphysical.
We run this test for comparison purposes only and do not recommend single star only models for use when analysing stellar populations.

\begin{figure}
    \centering
    \includegraphics[width=\columnwidth]{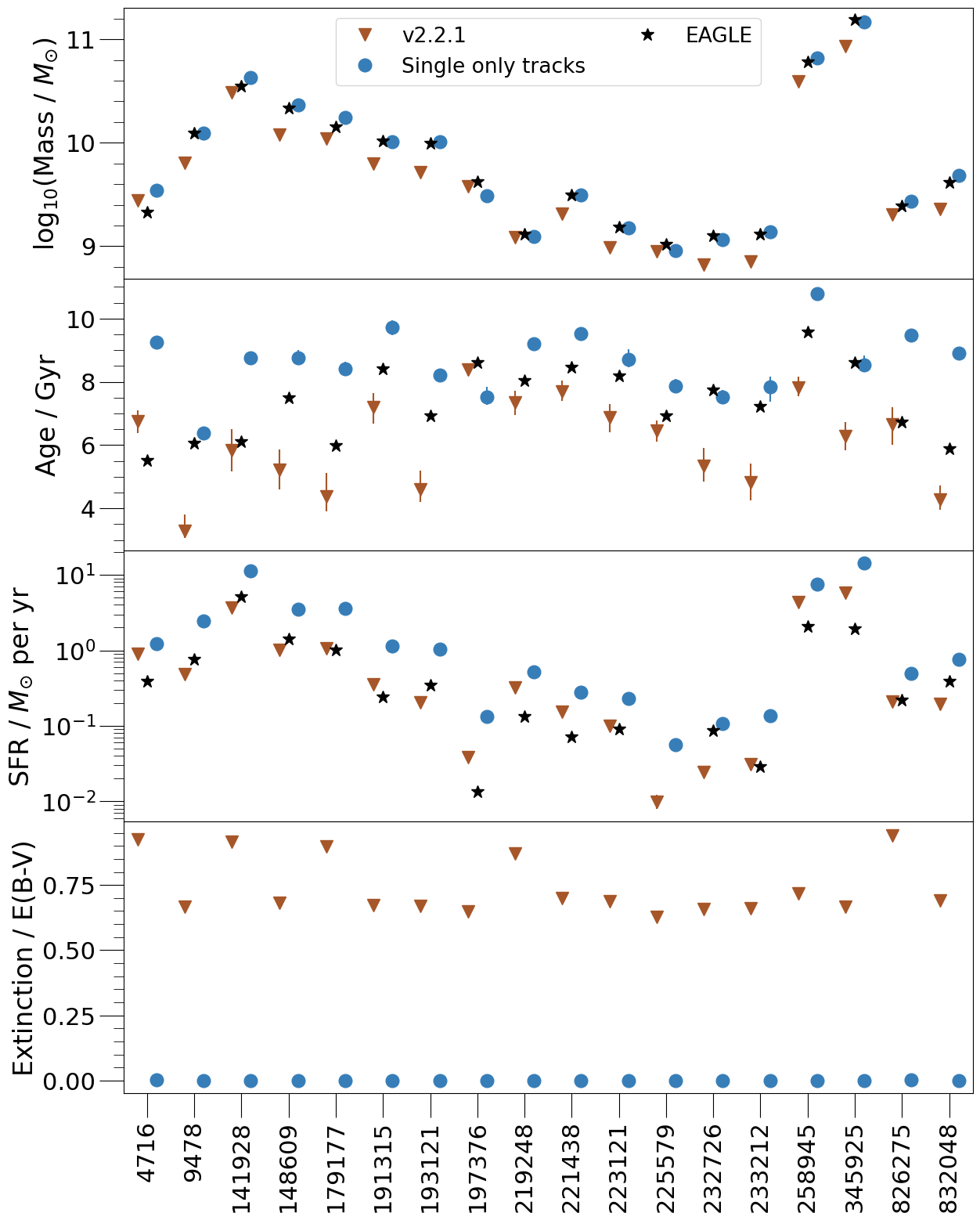} 
    \caption[]{Derived stellar mass, age, star formation rate, and extinction for each simulated galaxy, as inferred from spectral fits using SPS templates generated using different evolutionary tracks.
    The brown inverted triangles are the default tracks containing both single star and binary evolutionary tracks, while the blue circles use only the single star tracks.
    EAGLE catalogue values are shown as blacks stars.
    The derived extinction values for the single only tracks are non-zero but negligible ($E(B-V)<0.01$).}
    \label{fig:single_tracks_param_comp}
\end{figure}

\begin{figure}
    \centering
    \includegraphics[width=\columnwidth]{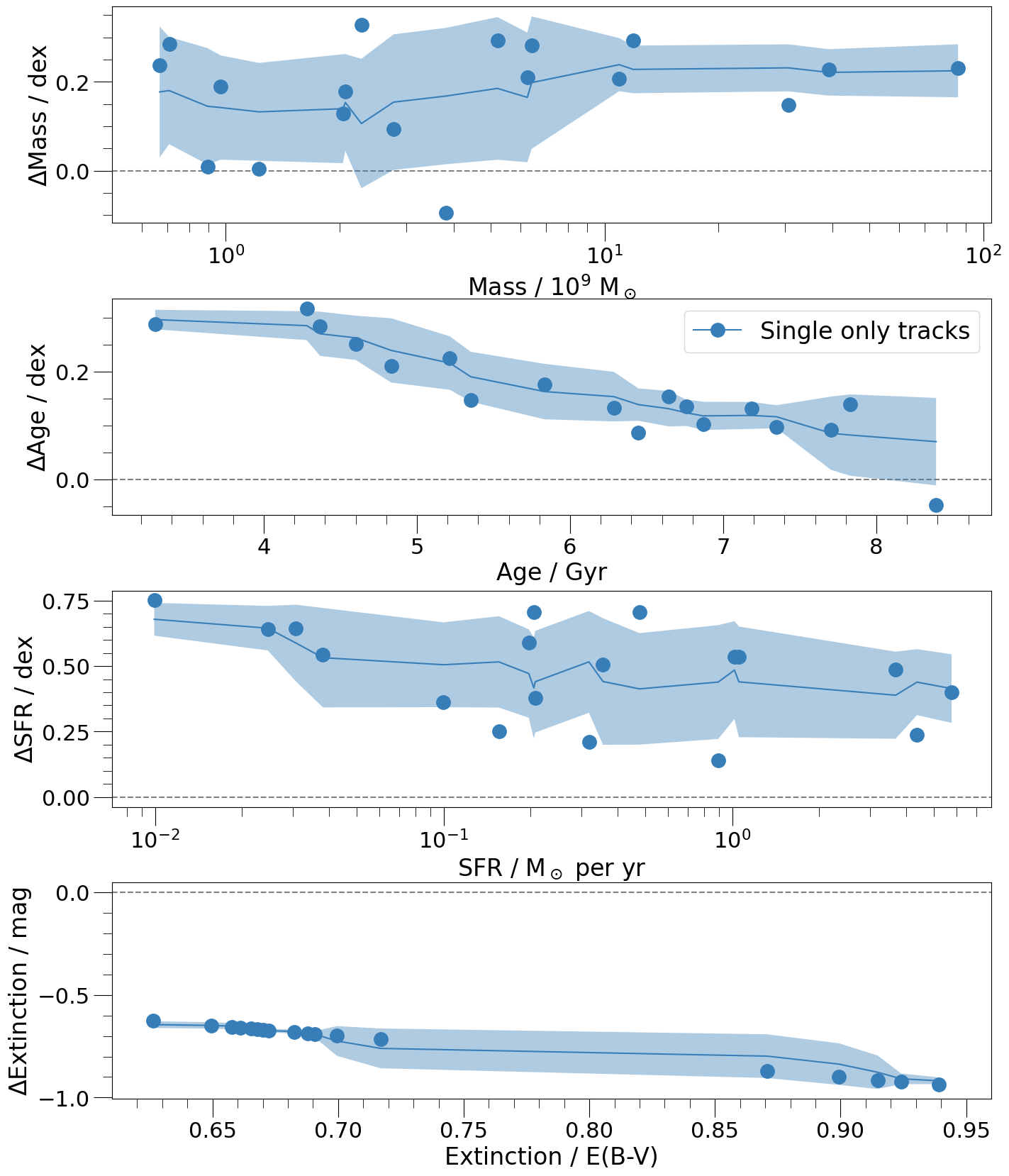} 
    \caption[]{Inferred galaxy property difference between the single only evolutionary track and the both single and binary incorporated track results, plotted as a function of the values obtained using the binary incorporated tracks.
    Each data point corresponds to a single simulated galaxy fit, while the solid line shows the running average over 6 data points with the shaded region the one-sigma uncertainty.
    Error ranges on individual galaxies are not included for clarity.}
    \label{fig:uncert_single_tracks_comp}
\end{figure}

Analysis follows the same methodology presented in Section~\ref{sec:tsc_meth}, using the same galaxy sample generated from the v2.2.1 BPASS models with default IMF and binary incorporated evolutionary tracks.
These are fit using \bagpipes\ and the same input parameters from Section~\ref{sec:tsc_bagpipes}, with one fit using the v2.2.1 default IMF, binary incorporated evolutionary tracks (still labelled v2.2.1 for consistency) and the other using v2.2.1 default IMF, single-star only evolutionary tracks (hereafter, single only tracks).
The inferred parameters for each simulated galaxy from both fits are shown in Figure~\ref{fig:single_tracks_param_comp} and the difference between the single only tracks and the binary incorporated tracks plotted in Figure~\ref{fig:uncert_single_tracks_comp}.
The mean offset in each derived parameter value is summarised in Table~\ref{tab:single_tracks_offsets}.

\begin{table*}
    \begin{center}
    \caption[]{Mean difference in inferred galaxy properties from the single star only evolutionary tracks results relative to those inferred using both single and binary evolutionary tracks.}
    \label{tab:single_tracks_offsets}
    \begin{tabular}{l | cccc}
        \hline
        Test & Stellar Mass / dex & Age / dex & SFR / dex & Extinction / mag \\
        \hline
        Single only tracks & $0.18 \pm 0.11$ & $0.16 \pm 0.09$ & $0.48 \pm 0.18$ & $-0.74 \pm 0.12$ \\
        \hline
    \end{tabular}
    \end{center}
\end{table*}

The derived parameters are significantly different as has been observed previously when comparing binary incorporated SPS models to models with only single-star tracks \citep{2022MNRAS.514.5706J}.
Binary interactions increase a star's lifetime, causing a given star to contribute for a longer time to the overall stellar population luminosity.
This increases the luminosity of the stellar spectrum relative to single-only tracks, resulting in a lower mass-to-luminosity ratio for binary SPS models.
Thus, to generate the same overall spectral luminosity, single-only SPS models will derive higher masses.
This analysis found the single only tracks derive increase masses by $0.18\pm0.11$\,dex.
This is in agreement with previous results, where \citet{2022MNRAS.514.5706J} derive $0.14\pm0.02$\,dex and \citet{2024ApJ...962...59O} derive $0.17\pm0.08$\,dex higher masses for single only track SPS models.

Increased SFRs for single only tracks follows from the same logic.
Binaries increasing the lifetime of stars results in a prolonged period over which hot, blue stars contribute to stellar spectra. 
This makes the stellar continuum in the spectra bluer, with increased emission in UV and blue-optical components, and increases the relative strength of spectral line features \citep[e.g.][]{2018MNRAS.477..904X}.
Thus, to reproduce this with single only tracks, a larger young, blue stellar population is required, deriving increased SFRs.
This analysis agrees with the SFRs derived in \citet{2022MNRAS.514.5706J}, which found single only track SPS models derive increased SFRs of $0.31\pm0.17$\,dex.

The most interesting parameters are the derived galaxy age and extinction.
Due to the need to increase the young stellar population to match the blue component and spectral line features, the overall continuum shape of the spectrum has been altered to be bluer.
Old stellar populations, as selected in the galaxy sample, have an overall redder continuum.
Thus, the spectrum needs to become redder.
This is achieved via an increased galaxy age, an increased extinction, or a combination of the two.
These two derived parameters are degenerate as they have the same overall effect on a spectrum.
In our results, the fits have applied the required reddening by increasing the age of the galaxy while applying negligible extinction ($E(B-V)<0.01$).
In nature, galaxies contain dust meaning these low extinction values are unphysical.

The inferred galaxy age and extinction parameters from the analysis in \citet{2022MNRAS.514.5706J} are significantly different to this work.
\citet{2022MNRAS.514.5706J} find no notable difference in derived age between the single only and binary incorporated tracks, while the derived extinction was $E(B-V)=0.064\pm0.022$\,mag higher for the single only tracks, opposite to the results from this paper.
They reported the increased extinction values were not just to redden the spectrum but also to match the infrared, dust emission part of the SED spectrum that they were also modelling.
This extra information from panchromatic SED fitting (extending into the far-infrared) helped break the degeneracy between age and extinction, whereby all the reddening was done through increased extinction rather than age.

Ultimately, this test has demonstrated two key facts.
Firstly, the uncertainties from SPS model component choices are larger than observational uncertainties no matter which model component is investigated, with similar derived uncertainties between the three components.
Secondly, full, panchromatic SED fitting involving both spectra and photometry should be used whenever the data is available, in order to break degeneracies and improve constrains.
This should be the case even if only the stellar or dust populations are being modelled.

\section{Validation of Spectra Generation Code} \label{sec:code_cal}
\begin{figure}
    \centering
    \includegraphics[width=\columnwidth]{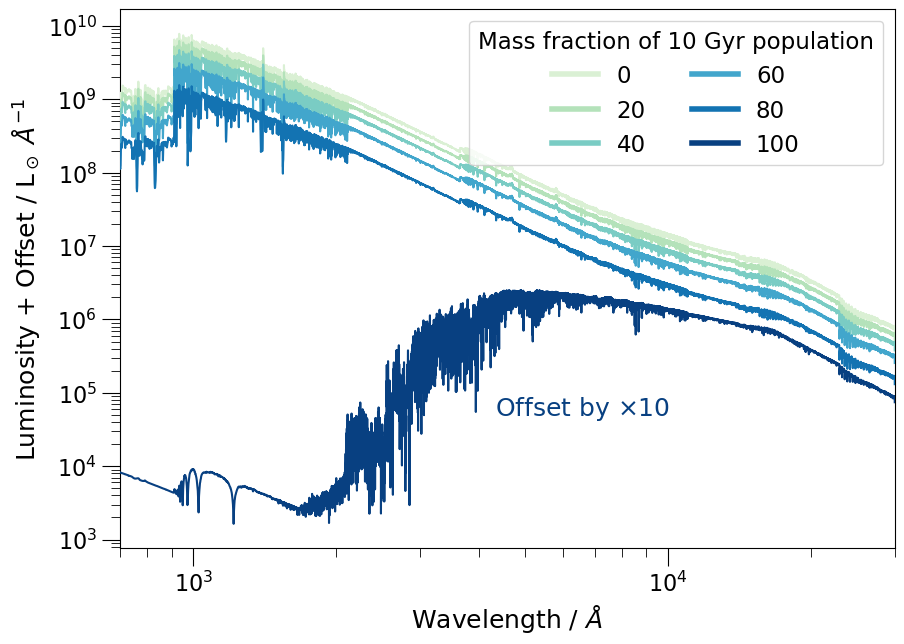} 
    \caption[]{Spectra of a $10^{10}$\,M$_\odot$ synthetic galaxy containing different mass fractions of 5\,Myr and 10\,Gyr stellar populations.
    The legend indicates the mass fraction in the 10\,Gyr stellar population, which increases from zero percent in the pale green spectrum (i.e. only 5\,Myr stars) to 100 percent in the dark blue spectrum (i.e. only 10\,Gyr stars).
    The flux for the dark blue spectrum has been increased by a factor of $10$.}
    \label{fig:synt_gals_cal_test}
\end{figure}

The code used to generate the mock spectra in this work, as described in Section~\ref{sec:spec_gen}, was validated through two tests.
The first involved constructing synthetic galaxies composed of a combination of 5\,Myr and 10\,Gyr star particles using different mass fractions of the two stellar populations.
This test ensures the resulting spectra reflect the expected characteristics of younger or older galaxies.
The results are presented in Fig~\ref{fig:synt_gals_cal_test}.
All spectra represent a $10^{10}$\,M$_\odot$ synthetic galaxy, with the fraction of mass in the 10\,Gyr stellar population increasing from the pale green line (entirely 5\,Myr stars) to the dark blue line (entirely 10\,Gyr stars), as indicated in the legend.

As expected, the 5\,Myr population dominates the spectra due to its much higher luminosity, outshining the 10\,Gyr population by several orders of magnitude.
In fact, the contribution from the 10\,Gyr stars becomes visible only when the 5\,Myr population is entirely absent.
Since luminosity scales as $L \propto M^{2.5}$ and the 5\,Myr population contains young, massive stars, it dominates the spectra.
These stars also give it the characteristic bluer spectrum, compared to the 10\,Gyr only synthetic galaxy which has lost these stars and thus appears redder.
Spectra containing both populations are normalised appropriately according to the mass fraction of the 5\,Myr population.
This test confirms that the code correctly generates spectra for galaxies composed of simple stellar populations, capturing the expected luminosity and colour evolution.

\begin{figure}
    \centering
    \includegraphics[width=\columnwidth]{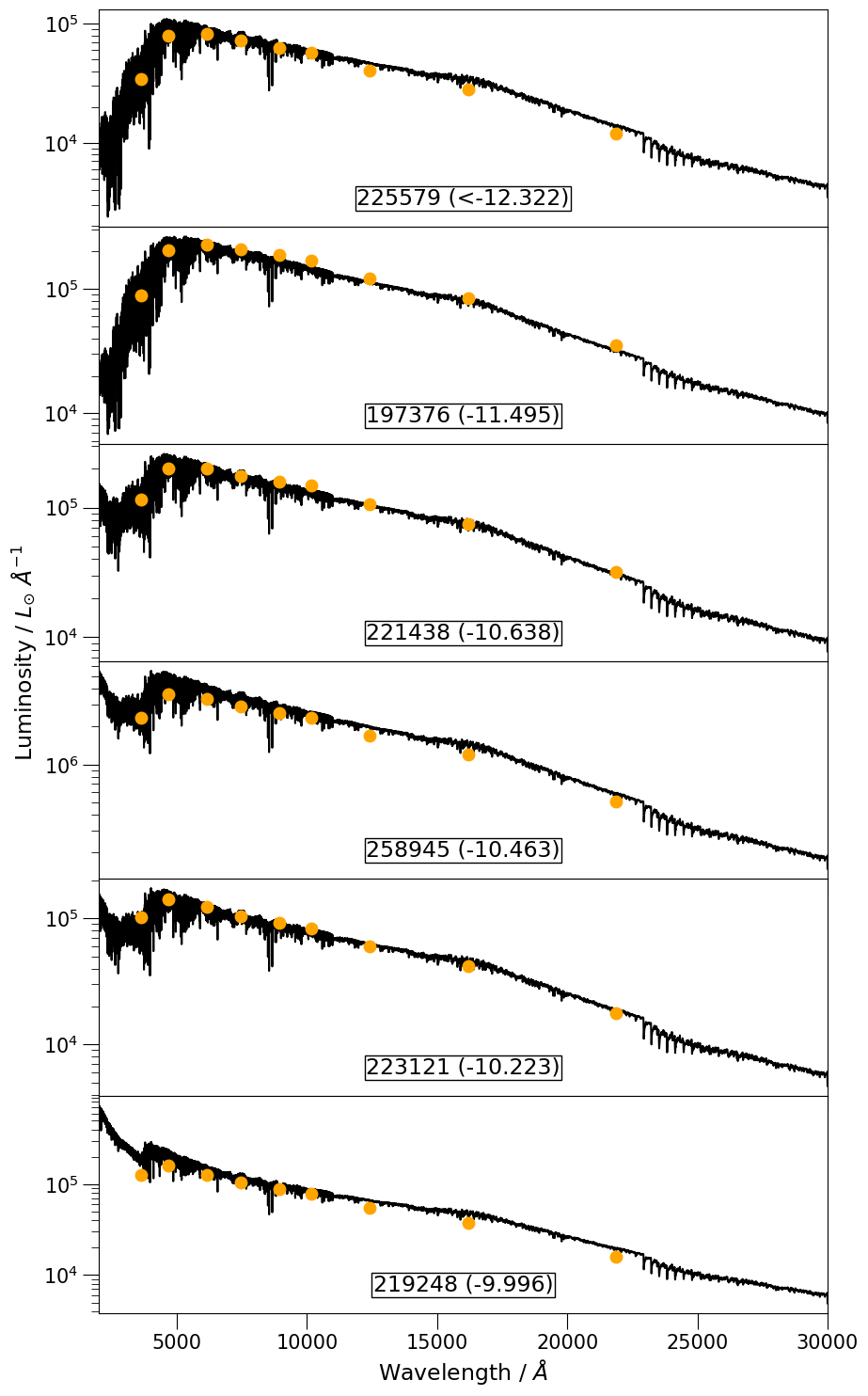} 
    \caption[]{Comparison of six mock spectra (in black) with photometry calculated by the EAGLE team overlaid as orange circles.
    The galaxy shown is labelled at the bottom of each plot, with the specific star formation rate given in brackets.
    Galaxy 225579 has an upper limit for the sSFR since its SFR in the catalogue is zero (see Table~\ref{tab:galaxy_ids}), calculated using the lowest SFR a gas particle has in the EAGLE simulation.}
    \label{fig:eagle_photo_comp}
\end{figure}

The second validation test used the complex stellar populations directly from the EAGLE simulations.
The EAGLE catalogue provides photometry for galaxies in the SDSS \textit{ugriz} \citep{2010AJ....139.1628D} and United Kingdom Infrared Telescope \textit{Y J H K} \citep[UKIRT,][]{2006MNRAS.367..454H} filter bands.
Photometry is calculated as non-dust-attenuated rest-frame broad-band magnitudes, measured within a 30\,physical-kpc spherical aperture using the \citet{2003MNRAS.344.1000B} SPS models, as detailed in \citet{2015MNRAS.452.2879T}.
We compare the catalogue photometry with the spectra generated by our code for six galaxies in Fig.~\ref{fig:eagle_photo_comp}.
Each panel indicates the galaxy ID and sSFR at the bottom.

For galaxy 225579, which has a reported SFR of zero (see Table~\ref{tab:galaxy_ids}), we estimate an upper limit for its sSFR using the lowest non-zero SFR value assigned to any gas particle in the EAGLE simulation ($\sim0.0005$\,M$_\odot$\,yr$^{-1}$).
Notably, the mock spectrum generated for galaxy 345925 overpredicts the photometry by approximately a factor of two.
This discrepancy arises because the galaxy has a large spatial extent such that a 30 physical-kpc aperture does not capture all of its star particles, whereas our mock spectra include these particles. 
This is the only galaxy in our sample where this aperture mismatch is expected to have a substantial impact.

The generated mock spectra agree well with the EAGLE photometry, reproducing the overall spectral shape and slopes.
Minor fluctuations do exist where the mock spectra can over- or under-predict the luminosity compared to the photometry, which arise due to the underlying SPS models used.
Our mock spectra use the BPASS models, while the photometry assume a \citet{2003MNRAS.344.1000B} template.
These models predict slightly different SSPs due to differences in assumed physics (e.g. inclusion of binaries) and assumed stellar spectral libraries.
Despite these differences, the mock spectra provide an excellent match to the catalogue photometry.

Furthermore, the quiescent galaxies identified by our classification (galaxies 225579 and 197376, top two panels of Fig.~\ref{fig:eagle_photo_comp}) display significantly redder spectra compared to the star-forming galaxies, with a steep luminosity drop-off at UV wavelengths.
This is consistent with their lack of young, massive stars due to halted star formation activity.
Overall, the close agreement between the mock spectra and EAGLE catalogue photometry validates our spectrum generation code, indicating that the method should not introduce significant discrepancies into the analysis.

\section{Assessing Impact of Dust Attenuation} \label{sec:impact_of_dust}
This study investigates how assumptions within SPS models influence the derivation of galaxy and stellar population properties.
Dust attenuation was included to mimic realistic, blind-fitting survey conditions where dust is a parameter to be constrained.
Other components (i.e. nebular emission) were excluded in our analysis since they are typically unconstrained in these surveys but nonetheless could introduce additional uncertainty.
In addition, dust not only alters the observed stellar emission but also introduces extra fitting parameters that may affect the derived results.
To assess the potential influence of dust on our findings, we conducted an additional test focusing exclusively on the intrinsic stellar emission by removing dust attenuation.

\begin{figure}
    \centering
    \includegraphics[width=\columnwidth]{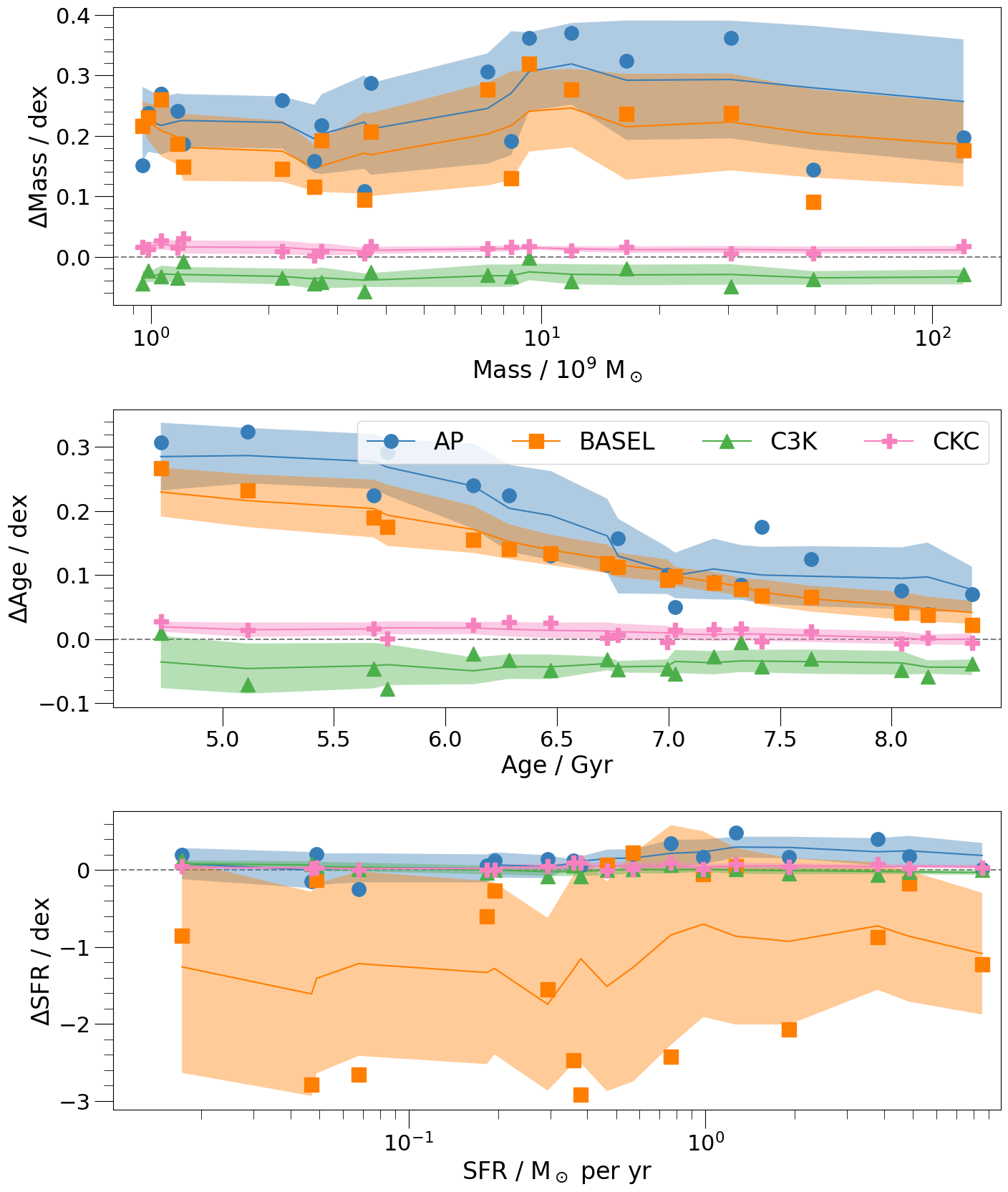} 
    \caption[]{Same as Fig.~\ref{fig:uncert_atmo_comp} but with no dust attenuation included either in the spectra generation or in the SED fitting procedure.}
    \label{fig:uncert_atmo_comp_no_dust}
\end{figure}

Mock spectra are generated for the same 18 galaxies following the same methodology as described in Section~\ref{sec:spec_gen}, but with the dust attenuation fixed to zero ($\mathrm{E(B-V)}=0$).
These zero-dust spectra were then fit with the same BPASS spectral library variants (Section~\ref{sec:tsc_spec_libs}) using \bagpipes\ and an identical fitting methodology (Section~\ref{sec:tsc_bagpipes}), except with dust attenuation fixed as zero in \bagpipes.
The results are shown in Fig.~\ref{fig:uncert_atmo_comp_no_dust}.
While on an individual galaxy basis there is a minor amount of scatter on any derived parameter, this variation is random, expected due to minor methodological differences, and negligible compared to variations introduced by SPS model choice.
Overall, the previously identified trends persist, with a $\sim0.2-0.3$\,dex offset in stellar mass, a decreasing offset in age with increasing age, and the BaSeL models predicting significantly lower SFRs by one to two orders of magnitude.
The only notable change is improved agreement between the AP models and the v2.2.1, CKC and C3K spectral models for the derived SFR.

These findings confirm that dust attenuation is not responsible for the variations observed in the derived parameters across different SPS model assumptions.
The high-quality spectra and photometry used in this study allow for tighter constraints on galaxy parameters, mitigating degeneracies that are more prominent in purely photometric analyses.
In typical survey conditions where only photometry is available, degeneracies between dust and other parameters could lead to a stronger impact from attenuation assumptions.
However, in this analysis, the magnitude and nature of the parameter offsets are primarily driven by differences between the SPS model assumptions, not by the inclusion of dust attenuation.

\section{Further Testing on IMF Variants } \label{sec:tsc_imf_test_append}

\begin{table*}
    \begin{center}
    \caption[]{Same as Table~\ref{tab:imf_offsets}, but for the results where mock spectra generated from different IMF prescriptions are all fit using a single SPS template (v2.2.1 with the default IMF).}
    \label{tab:imf_offsets_test}
    \begin{tabular}{l | cccc}
        \hline
        Model & Stellar Mass / dex & Age / dex & SFR / dex & Extinction / mag \\
        \hline
        Shallow & $-0.16 \pm 0.07$ & $-0.03 \pm 0.04$ & $-0.03 \pm 0.13$ & $0.11 \pm 0.12$ \\
        Steep & $0.06 \pm 0.06$ & $0.04 \pm 0.05$ & $-0.12 \pm 0.08$ & $-0.05 \pm 0.10$ \\
        Continuous & $-0.11 \pm 0.05$ & $-0.01 \pm 0.05$ & $-0.12 \pm 0.03$ & $0.03 \pm 0.05$ \\
        C03 & $0.01 \pm 0.06$ & $-0.01 \pm 0.04$ & $0.03 \pm 0.08$ & $0.04 \pm 0.09$ \\
        \hline
    \end{tabular}
    \end{center}
\end{table*}

\begin{figure}
    \centering
    \includegraphics[width=\columnwidth]{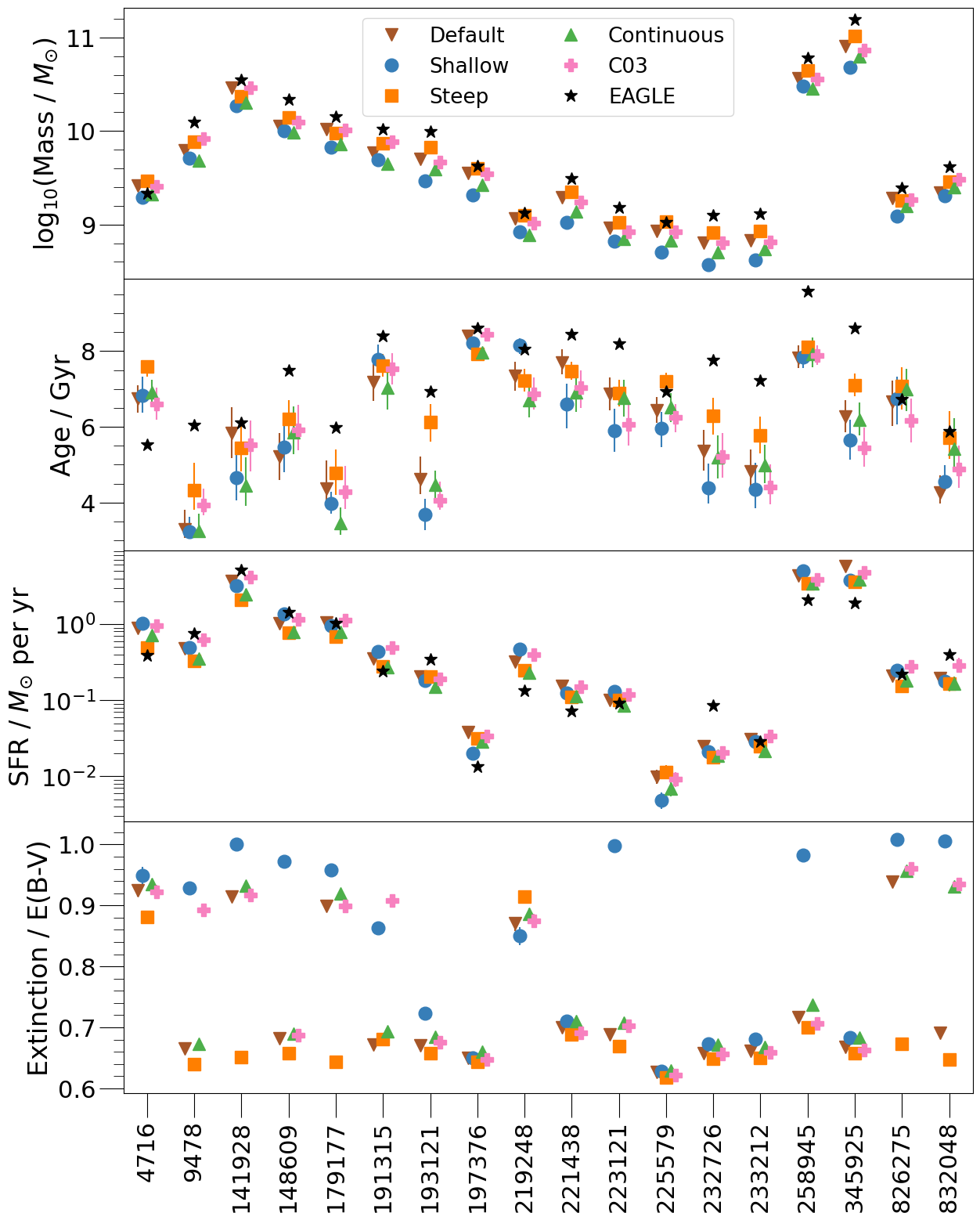} 
    \caption[]{Same as Fig.~\ref{fig:imf_param_comp}, but for the results where mock spectra generated from different IMF prescriptions are all fit using a single SPS template (v2.2.1 with the default IMF).}
    \label{fig:imf_param_comp_test}
\end{figure}

\begin{figure}
    \centering
    \includegraphics[width=\columnwidth]{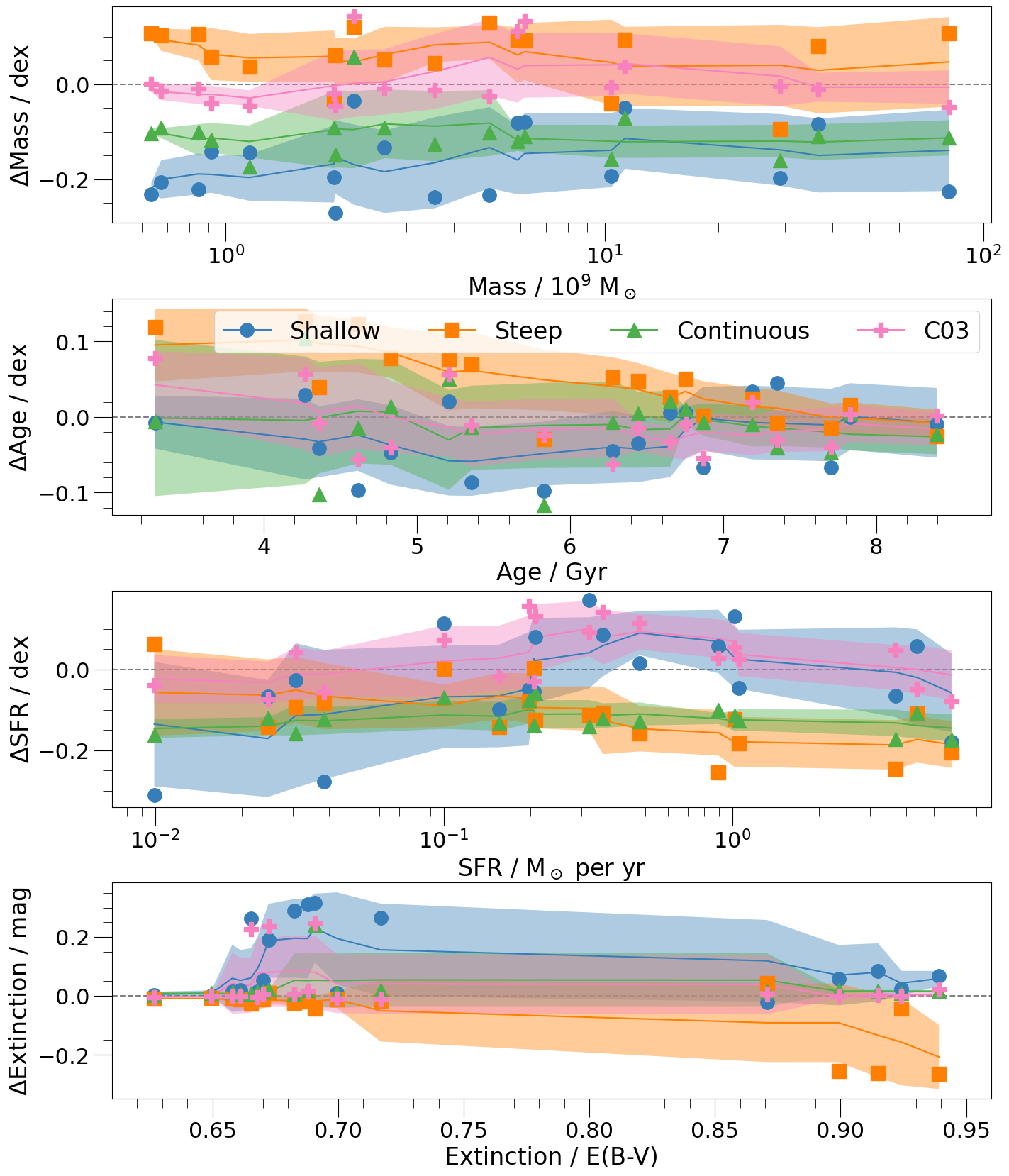} 
    \caption[]{Same as Fig.~\ref{fig:uncert_imf_comp}, but for the results where mock spectra generated from different IMF prescriptions are all fit using a single SPS template (v2.2.1 with the default IMF).}
    \label{fig:uncert_imf_comp_test}
\end{figure}

To verify the findings from our analysis, we reassess the uncertainties cause from the IMF prescription applying a altered version of our methodology involving two key modifications: a mock galaxy spectrum is generated for each IMF prescription rather than solely for the v2.2.1 default framework, and all spectra are only fit using only a single SPS template (v2.2.1 framework with the default IMF).
All other input priors remain unchanged.
The expected outcome is that the direction of the parameter offsets relative to the default IMF should be inverted.
For instance, if fitting a galaxy spectrum generated with a shallow IMF prescription using a steep IMF template leads to an overestimated SFR (relative to default), then the reverse configuration should result in an underestimated SFR.

The derived parameters from each fit are plotted in Fig.~\ref{fig:imf_param_comp_test}.
Differences relative to the default IMF framework are shown in Fig.~\ref{fig:uncert_imf_comp_test}, plotted as a function of the default IMF-derived parameter values.
The mean difference between default IMF prescription and the others is given in Table~\ref{tab:imf_offsets_test}.
Age and extinction values are generally consistent across all IMF variants.
In contrast, stellar mass and SFR exhibit more significant variation.
As expected, these show the opposite trend to before, although with slightly better agreement to the default values than was found in Section~\ref{sec:imf_everything}.

The improved agreement observed in this test is likely due to increased uncertainties in this new methodology, primarily stemming from the treatment of dust attenuation.
For each simulated EAGLE galaxy, all mock spectra are generated using the same \citet{2018ApJ...859...11S} dust attenuation law (star-forming or quiescent), determined by the galaxy's sSFR in the EAGLE catalogue.
However, unlike in the earlier methodology, the fitting process may apply different attenuation laws to mock spectra produced from the same simulated galaxy but generated using various IMF assumptions.
This is because classification when fitting into star-forming or quiescent is based on the colours of each spectrum, which in turn depend on the chosen SPS variant.
Top-heavy IMFs, with a greater proportion of high-mass stars, produce bluer spectra that are more likely to be classified as star-forming.
Conversely, bottom-heavy IMFs generate redder spectra more typical of quiescent galaxies.
This trend is evident in the bottom panel of Fig.~\ref{fig:imf_param_comp_test}, where shallower IMFs imply higher extinction values due to the application of the star-forming dust attenuation law.
The inconsistency in attenuation priors between mock spectra for the same simulate galaxy introduces additional scatter in the parameter trends, resulting in larger uncertainties.

Overall, this test demonstrates two key points.
First, after accounting for the increased scatter, the direction of derived parameter differences is generally reversed compared to the results in Section~\ref{sec:imf_everything}.
This confirms that uncertainties arising from our results reflect that caused from SPS assumption variations, while relative parameter value differences reflect the application of the assumptions to any sample.
Second, assuming specific SPS assumptions for galaxy emission can significantly influence the interpretation of galaxy properties, particularly the choice of dust attenuation law and the classification of galaxies as star-forming or quiescent.
This sensitivity to attenuation law selection has been shown in previous studies to affect the derived stellar masses and SFRs of galaxies \citep[e.g.][]{2013ApJ...775L..16K, 2022ApJ...931...14L}.


\bsp	
\label{lastpage}
\end{document}